\documentclass[3p]{elsarticle}

\usepackage{lineno,hyperref}
\modulolinenumbers[5]

\journal{Annals of Physics}

\usepackage{graphicx}
\usepackage{bm}

\begin{document}

\begin{frontmatter}

\title{Quantum theory as the most robust description of reproducible experiments}


\author{Hans De Raedt}
\address{Department of Applied Physics, Zernike Institute for Advanced Materials,\\
University of Groningen, Nijenborgh 4, NL-9747AG Groningen, The Netherlands}
\ead{h.a.de.raedt@rug.nl}
\author{Mikhail I. Katsnelson}
\address{Radboud University Nijmegen, Institute for Molecules and Materials,
Heyendaalseweg 135, NL-6525AJ Nijmegen, The Netherlands}
\ead{M.Katsnelson@science.ru.nl}
\author{Kristel Michielsen\fnref{myfootnote}}
\address{Institute for Advanced Simulation, J\"ulich Supercomputing Centre,\\
Forschungszentrum J\"ulich, D-52425 J\"ulich, Germany\\and\\
RWTH Aachen University, D-52056 Aachen, Germany}
\fntext[myfootnote]{Corresponding author}
\ead{k.michielsen@fz-juelich.de}

\begin{abstract}
It is shown that the basic equations of quantum theory
can be obtained from a straightforward application of logical inference
to experiments for which there is uncertainty about individual events
and for which the frequencies of the observed events
are robust with respect to small changes in the conditions
under which the experiments are carried out.
\end{abstract}


\end{frontmatter}

\linenumbers

\section{Introduction}
Quantum theory has proven to be extraordinary powerful to describe
a vast amount of very different experiments in
(sub)-atomic, molecular and condensed matter physics, quantum optics and so on.
Remarkably, after so many extremely successful practical applications,
there are still hot debates about conceptual backgrounds of quantum theory,
and attempts to clarify the success continue until now.

The success of quantum theory reminds us of an example of another very successful theory, namely classical thermodynamics.
Einstein said: ``Classical thermodynamics is the only physical theory of universal content which I am convinced
will never be overthrown, within the framework of applicability of its basic concepts''~\cite{EINS79}.
Can we say that we understand the reasons of this success from the point of view of a more fundamental theory?
Strictly speaking, a rigorous derivation of, say, the second law of thermodynamics
from classical (or quantum) mechanics is lacking and therefore the answer should be ``no''
but in practice this does not matter too much.
Our belief in thermodynamics is not based on mathematical deduction but on
its power to account for everyday experience.

It has been emphasized many times that our description of physical phenomena
at some level of observation is essentially independent of our view of ``underlying'' levels~\cite{LAUG05}.
In the present paper, we apply the same world view to nonrelativistic quantum theory.
Adopting this view immediately distinguishes our line of thinking from
approaches that assume an underlying ontology~\cite{BOHM52,PENA96,THOO97,PENA05,THOO07}
or formulate quantum theory starting from various sets of axioms~\cite{%
FRIE89,VSTO95,REGI98,HALL00,HARD01,LUO02,FRIE04,BUB07,HARD07,CAVE07,PALG08,KHRE09,KAPS10,CHIR10,CHIR11,MASA11,BRUK11,%
KHRE11,KHRE11a,SKAL11,KAPS11,ORES12,SANT12,KLEI12,KLEI12a,FLEG12,KAPU13,FUCH13,HOLI14}.
We start with something that is as reliable as one can imagine, which in our view,
are the principles of logical inference~\cite{COX46,COX61,TRIB69,SMIT89,JAYN03} (a brief, formal introduction is given below) 
and ask the question: what should be added to these principles in order to derive,
for instance, the (nonrelativistic) Schr\"odinger equation?
The answer is that it suffices to add Bohr's correspondence principle in a probabilistic sense. 

The present work explores the possibility of exploiting
logical inference~\cite{COX46,COX61,TRIB69,SMIT89,JAYN03} that is inductive reasoning
to give a rational explanation for the success
of quantum theory as a description of a vast class of physical phenomena.
We are not concerned with the various interpretations~\cite{KHRE09,HOME97,BALL01,GRIF02}
of quantum theory.

We introduce the basic ideas of our approach by starting with
a few quotes of Niels Bohr:
\begin{enumerate}
\item
There is no quantum world.
There is only an abstract physical description.
It is wrong to think that the task of physics is to find out how nature {\it is}.
Physics concerns what we can {\it say} about nature~\cite{PETE63}.
\item
Physics is to be regarded not so much as the study of something a priori given,
but rather as the development of methods of ordering and surveying human experience.
In this respect our task must be to account for such experience in a manner independent
of individual subjective judgment and therefore objective in the sense that it can be
unambiguously communicated in ordinary human language~\cite{BOHR99}.
\item
The physical content of quantum mechanics is exhausted by its power to formulate
statistical laws governing observations under conditions specified in plain language~\cite{BOHR99}.
\end{enumerate}
The first two sentences of the first quote may be read as a suggestion to dispose of,
in Mermin's words~\cite{MERM09}, the ``bad habit'' to
take mathematical abstractions as the reality of the events (in the everyday sense of the word)
that we experience through our senses.
Although widely circulated, these sentences are reported by Petersen~\cite{PETE63}
and there is doubt that Bohr actually used this wording~\cite{PLOT10a}.
The last two sentences of the first quote and the second quote suggest that we should try to describe
human experiences (confined to the realm of scientific inquiry)
in a manner and language which is unambiguous and independent of the individual subjective judgment.
Of course, the latter should not be construed to imply that
the observed phenomena are independent of the choices made
by the individual(s) in performing the scientific experiment~\cite{SCHO07}.

The third quote suggests that quantum theory is a powerful
language to describe a certain class of statistical experiments
but remains vague about the properties of the class.
Similar views were expressed by other fathers of quantum mechanics, e.g.,
Max Born and Wolfgang Pauli~\cite{LAUR97}.
They can be summarized as ``Quantum theory describes our {\it knowledge} of the atomic
phenomena rather than the atomic phenomena themselves''.
Our aim is, in a sense, to replace the philosophical components of these statements
by well-defined mathematical concepts and to carefully study their relevance for physical phenomena.
Specifically, by applying the general formalism of logical inference
to a well-defined class of statistical experiments,
the present paper shows that quantum theory is indeed the kind of language
envisaged by Bohr.

Theories such as Newtonian mechanics, Maxwell's electrodynamics,
and Einstein's (general) relativity are deductive in character.
Starting from a few axioms, abstracted from experimental observations
and additional assumptions about the irrelevance of a large number
of factors for the description of the phenomena of interest,
deductive reasoning is used to prove or disprove
unambiguous statements, propositions,
about the mathematical objects which appear in the theory.

The method of deductive reasoning conforms to the Boolean algebra of propositions.
The deductive, reductionist methodology has the appealing
feature that one can be sure that the propositions are either
right or wrong, and disregarding the possibility that some
of the premises on which the deduction is built may not
apply, there is no doubt that the conclusions are correct.
Clearly, these theories successfully describe a wide range of physical phenomena
in a manner and language which is unambiguous and independent of the individual.

At the same time, the construction of a physical theory, and a scientific theory in general, from ``first principles''
is, for sure, not something self-evident, and not even safe.
Our basic knowledge always starts from the middle, that is, from the world of macroscopic objects.
According to Bohr, the quantum theoretical description crucially depends on the existence of macroscopic objects which can be used as
measuring devices. For an extensive analysis of the quantum measurement process from a dynamical point of view see Ref.~\cite{NIEU13}.
Most importantly, the description of the macroscopic level is robust, that is, essentially independent of the
underlying ``more fundamental'' picture~\cite{LAUG05}.
As will be seen later, formalizing the notion of ``robustness'' is key to derive the basic equations of quantum theory
from the general framework of logical inference.

Key assumptions of the deductive approach are
that the mathematical description is a complete
description of the experiment under consideration
and that there is no uncertainty
about the conditions under which the experiment is carried out.
If the theory does not fully account for all the relevant aspects
of the phenomenon that we wish to describe,
the general rules by which we deduce whether a proposition is true or false can no longer be used.
However, in these circumstances, we can still resort to logical inference~\cite{COX46,COX61,TRIB69,SMIT89,JAYN03}
to find useful answers to unambiguous questions.
Of course, in general it will no longer be possible to say whether
a proposition is true or false, hence there will
always remain a residue of doubt.
However, as will be shown, the description obtained through logical inference
may also be unambiguous and independent of the individual.

In the present paper, we demonstrate that the basic equations of
quantum theory directly follow from logical inference
applied to experiments in which there is
\begin{enumerate}[(i)]
\item
uncertainty about individual events,
\item
the stringent condition that certain properties of the collection of events are reproducible,
meaning that they are robust with respect to small changes in the conditions
under which the experiments are carried out.
\end{enumerate}
It is the latter that renders the theoretical description
unambiguous and independent of the individual.
In addition, our work provides a rational foundation for Bohr's philosophical viewpoints
embodied in quotes (1--3).

The paper is structured as follows.
Section~\ref{sec2} contains a brief introduction to the algebra of logical inference~\cite{COX46,COX61,TRIB69,SMIT89,JAYN03},
a mathematical framework which formalizes
the patterns of plausible reasoning exposed by P\'olya~\cite{POLY54}.
This mathematically precise formalism expresses what most people would consider to be rational reasoning.
The key concept is the notion of the plausibility that a proposition is true
given that another proposition is true.
Section~\ref{sec3} discusses the role of uncertainties in experiments and classifies their theoretical
descriptions.
In Section~\ref{EPRB}, we show in detail how logical inference can be used
to derive the quantum theoretical description of the Einstein-Podolsky-Rosen-Bohm
thought experiment without invoking a single concept of quantum theory.
Section~\ref{SG} uses the Stern-Gerlach experiment to illustrate how the approach of Section~\ref{sec2}
may be extended by adding features abstracted from the experiment.
These two sections are based on earlier attempts to derive
the expressions of quantum theory by logical inference~\cite{RAED06a,RAED07c}.
Finally, we demonstrate that the time-independent (Section~\ref{SE})
and time-dependent (Section~\ref{TDSE}) Schr\"odinger equation
can be derived by logical inference from the assumption
that the experiment yields reproducible data.
A discussion of general aspects of our approach and
conclusions are given in Section~\ref{CONC}.

\section{The algebra of logical inference}\label{sec2}

Obviously, any attempt to capture the process of human reasoning
by which the events are registered by our senses and are brought in relation to each other,
leading to abstract concepts, is bound to create more problems than we can solve at this time.
However, if we are only concerned about quantifying the truth of a proposition
given the truth of another proposition, it is possible to construct
a mathematical framework, an extension of Boolean logic,
that allows us to reason in a manner which is unambiguous and independent of the
individual, in particular if there are elements of uncertainty in the description~\cite{COX46,COX61,TRIB69,SMIT89,JAYN03}.

In this section, we briefly introduce the concepts that are necessary for the purpose of the present paper.
For a detailed discussion of the foundations of
plausible reasoning, its relation to Boolean logic and the derivation of the
rules of logical inference, the reader is advised to consult the papers~\cite{COX46,SMIT89} and books~\cite{COX61,TRIB69,JAYN03}
from which our summary has been extracted.

We start by listing three so-called ``desiderata'' from which
the algebra of logical inference can be derived~\cite{COX61,TRIB69,SMIT89,JAYN03}.
The formulation which follows is taken from Ref.~\cite{SMIT89}.
\begin{enumerate}[{\bf Desideratum} \bf 1.]
\item
{\it Plausibilities are represented by real numbers.}
The plausibility that a proposition $A$
is true conditional on proposition $B$ being true
will be denoted by $P(A|B)$.
\item
{\it Plausibilities must exhibit agreement with rationality.}
As more and more evidence supporting
the truth of a proposition becomes available,
the plausibility should increase monotonically and continuously
and the plausibility of the negation of the proposition
should decrease monotonically and continuously.
Moreover, in the limiting cases that the proposition $A$ is known to be either true
or false, the plausibility $P(A|B)$ should conform to the rules of deductive reasoning.
In other words, plausibilities must be in qualitative agreement with
the patterns of plausible reasoning uncovered by P\'olya~\cite{POLY54}.
\item
{\it All rules relating plausibilities must be consistent.}
Consistency of rational reasoning demands that if the rules
of logical inference allow a plausibility to be obtained in more than one way,
the result should not depend on the particular sequence of operations.
\end{enumerate}
These three desiderata only describe the essential features of the plausibilities
and definitely do not constitute a set of axioms which plausibilities have to satisfy.
It is a most remarkable fact that these desiderata suffice to uniquely determine
the set of rules by which plausibilities may be manipulated~\cite{COX61,TRIB69,SMIT89,JAYN03}.

Omitting the derivation, it follows that plausibilities
may be chosen to take numerical values in the range $[0,1]$ and
obey the rules~\cite{COX61,TRIB69,SMIT89,JAYN03}
\begin{enumerate}[a.]
\item
$P(A|Z)+P({\bar A}|Z)=1$ where
${\bar A}$ denotes the negation of proposition $A$
and $Z$ is a proposition assumed to be true.
\item
$P(AB|Z)=P(A|BZ)P(B|Z)=P(B|AZ)P(A|Z)$ where
the ``product'' $BZ$ denotes the logical product (conjunction)
of the propositions $B$ and $Z$,
that is the proposition $BZ$ is true if both $B$ and $Z$ are true.
This rule will be referred to as ``product rule''.
It should be mentioned here that it is not allowed
to define a plausibility for a proposition
conditional on the conjunction of mutual exclusive
propositions. Reasoning on the basis of two of more contradictory premises
is out of the scope of the present paper.
\item
$P(A{\bar A}|Z)=0$ and $P(A+{\bar A}|Z)=1$
where the ``sum'' $A+B$ denotes the logical sum (inclusive disjunction)
of the propositions $A$ and $B$,
that is the proposition $A+B$ is true if either $A$ or $B$ or both are true.
These two rules show that
Boolean algebra is contained in the algebra of plausibilities.
\end{enumerate}
The rules (a--c) are unique. Any
other rule which applies to plausibilities represented by real numbers
and is in conflict with rules (a--c)
will be at odds with rational reasoning and consistency~\cite{TRIB69,SMIT89,JAYN03}.

The reader will no doubt recognize that rules (a--c) are identical
to the rules by which we manipulate probabilities~\cite{JAYN03,KEYN21,FELL68,GRIM95}.
However, the rules (a--c) were not postulated.
They were derived from general considerations about
rational reasoning and consistency only.
Moreover, concepts such as sample spaces,
probability measures etc., which are an essential part of the mathematical
foundation of probability theory~\cite{FELL68,GRIM95}, play no role
in the derivation of rules (a--c).
In fact, if Kolmogorov's axiomatic
formulation of probability theory would have
been in conflict with rules (a--c), we believe
that this formulation would long have been disposed of because
it would yield results which are in conflict with rational reasoning.
Perhaps most important in the context of quantum theory
is that in the logical inference approach uncertainty about an event does not
imply that this event can be represented by a random variable
as defined in probability theory~\cite{GRIM95}.

We emphasize that there is a significant conceptual difference
between ``mathematical'' probabilities and plausibilities.
Mathematical probabilities are elements of an axiomatic framework
which complies with the algebra of logical inference.
Plausibilities are elements of a language
which also complies with the algebra of logical inference
and serve to facilitate communication,
in an unambiguous and consistent manner,
about phenomena in which there is uncertainty.

The plausibility $P(A|B)$ is an intermediate mental construct that
serves to carry out inductive logic, that is rational reasoning,
in a mathematically well-defined manner~\cite{TRIB69}.
Plausibilities are concepts resulting from human reasoning
about observed events and their relationships but are not
the ``cause'' of these events.
In general, $P(A|B)$ may express the degree of belief of an individual that
proposition $A$ is true, given that proposition $B$ is true.
However, in the present paper, we explicitly exclude applications of this
kind because they do not comply with our main goal,
namely to describe phenomena ``in a manner independent
of individual subjective judgment'', see Bohr's quote (2).

\begin{center}
\framebox{
\parbox[t]{0.9\hsize}{%
To take away this subjective connotation of the word ``plausibility'',
from now on
we will simply call $P(A|B)$ the ``inference-probability'' or ``i-prob'' for short.
}
}
\end{center}

The algebra of logical inference is the foundation for
powerful tools such as the maximum entropy method
and Bayesian analysis~\cite{TRIB69,JAYN03}.
Although not formulated in the language of logical inference used in the present paper,
Jaynes' papers on the relation between information and (quantum) statistical mechanics~\cite{JAYN57a,JAYN57b}
are perhaps the first to ``derive'' theoretical descriptions
using this general methodology of scientific reasoning.
As we show in this paper, quantum theory also derives
from the application of the algebra of logical inference.

It is important to keep in mind that
the rules of logical inference are not bound by ``the laws of physics''.
In particular, logical inference also applies to situations where
there are no causal relations between the events~\cite{TRIB69,JAYN03}.
The point of view taken in this paper is that the laws of physics should provide a consistent description of
relations between certain events that
we perceive by our senses and therefore they should conform to the rules of logical inference.
Although extracting cause-and-effect relationships from empirical evidence
by rational reasoning
should follow the rules of logical inference, in general the latter cannot
be used to establish cause-and-effect relationships~\cite{JAYN03,PEAR09,PLOT11b}.

A comment on the notation used throughout this paper is in order.
To simplify the presentation, we make no distinction between an event such as ``detector D has fired''
and the corresponding proposition ``$D =$ detector D has fired''.
If we have two detectors, say $D_x$ where $x=\pm1$, we write
$P(x|Z)$ to denote the i-prob
of the proposition
that detector $D_x$
fires, given that proposition $Z$ is true.
Similarly, the i-prob of the proposition
that two detectors $D_x$ and $D_y$ fire, given that proposition $Z$ is true,
is denoted by $P(x,y|Z)$.
Obviously, this notation generalizes to more than two propositions.

\section{Quantum theory as an instance of logical inference}\label{sec3}

The theoretical description of ``classical physics''
applies to phenomena for which there is absolute certainty about the outcome of
each individual experiment on each individual object~\cite{PLOT10,PLOT11,PLOT12}.
In mapping
the experimental data which are necessarily represented by a limited number of bits, that is by integers,
onto the theoretical abstractions in terms of real numbers,
it is assumed that the necessarily finite precision of the experiment
can be increased without limit, at least in principle, and that there is
a one-to-one mapping between the values of the variables in the theory
and the values of the corresponding quantities measured in experiment.

In real experiments there is always uncertainty about some
factors which may or may not influence the outcome of the measurements.
In the realm of classical physics, standard techniques of statistical analysis
are used to deal with this issue.
It is postulated that these ``imperfections'' in the experimental data
are not of fundamental importance but are technical
in nature and can therefore be eliminated, at least in principle~\cite{PLOT10,PLOT11,PLOT12}.

Quantum theory is fundamentally different
from classical theories in that there may be uncertainties about each individual event,
uncertainties which cannot be eliminated, not even in principle~\cite{PLOT10,PLOT11,PLOT12}.
Clearly, this is a statement about the theory, not about the observed phenomena themselves.
The outcome of a real experiment, be it on ``classical'' or ``quantum'' objects,
is always subject to uncertainties in the conditions under which the experiment is carried out.
However, this issue is not of direct concern to us here because we
only want to explore whether the quantum theoretical description, not the phenomena themselves, can be derived
from logical inference applied to certain thought experiments.

Summarizing, we may classify theoretical abstractions of scientific experiments as follows:
\begin{enumerate}[\bf {Category} 1.]
\item
The conditions under which the experiment is carried out are known and fixed
for the duration of the experiment and there is no uncertainty about each event.
\item
Each event under known conditions is certain
but the conditions under which the experiment is carried out may be uncertain.
\item
There may be uncertainty about each event
and the conditions under which the experiment is carried may be uncertain.
\end{enumerate}
A laboratory experiment always falls in category 3.
In a strict sense, numerical experiments on a digital computer belong to category 1.
However, disregarding the fact that in the course of the numerical experiment
the time-evolution of each individual bit of the computer is completely determined and known,
in practice, the complexity of the numerical simulation is often so large that
the variables of interest may exhibit behavior that
is similar to the one observed in experiments belonging to category 2 and 3.

In the theoretical description of a real experiment,
it makes sense to simplify matters by first exploring
models that belong to category 1 (classical physics)
and if no satisfactory description is obtained
to consider models of category 2 (classical physics supplemented with probability theory).
If the latter fails to describe the experiment too,
we can still try models in category 3.

The fact that laboratory experiments always belong to category 3 has an important implication.
A basic requirement for any scientific experiment is that
the analysis of the data yields quantities (e.g. frequencies, averages, correlations, etc.)
that exhibit a high degree of reproducibility.
Only then it may make sense to attempt drawing scientifically meaningful conclusions from these data.
Clearly, this requirement restricts the uncertainties on the conditions
under which the experiment is carried out.
If these uncertainties fluctuate wildly with each measurement, it
is unreasonable to expect reproducible results.

Therefore, it seems justified to limit attention to a subset of theoretical models of category 3
which satisfies the following criteria:
\begin{enumerate}[\bf {Category} 3a.]
\item
There may be uncertainty about each event.
The conditions under which the experiment is carried out may be uncertain.
The frequencies with which events are observed are reproducible
and robust against small changes in the conditions.
\end{enumerate}
As we show in this paper, the rules of logical inference applied to
models belonging to category 3a rather straightforwardly lead
to the basic equations of quantum theory.
The derivation has a generic structure.
The first step is to list the features of the experiment that are deemed
to be relevant and to introduce the i-probs of the individual events.
The second step is to impose the condition that the experiment
yields reproducible results, not on the level of individual events,
but on the level of averages of many events.
The result of the second step is a functional of the i-prob,
the minimum of which yields an expression for the i-prob
which is identical to the corresponding probability obtained
from the quantum theoretical description of the experiment.

\section{Einstein-Podolsky-Rosen-Bohm thought experiment}\label{EPRB}

As a first illustration, we consider Bohm's version of
the Einstein-Podolsky-Rosen thought experiment~\cite{EPR35,BOHM51}.
To head off possible misunderstandings, the derivation presented
in this section does not add anything to the ongoing discussions
about locality, realism, etc.
in relation to the violation of
Bell-like inequalities~\cite{RAED07c,PENA72,FINE74,FINE82,BROD93,KUPC86,JAYN89,SICA99,HESS01a,HESS05,KRAC05,SANT05,NIEU09,KARL09,NIEU11,RAED11a}.

We choose the Einstein-Podolsky-Rosen-Bohm (EPRB) thought experiment as the first example
because it seems to be the simplest nontrivial model for
demonstrating how the logical inference approach works.
Indeed, a straightforward application
of the ideas of Section~\ref{sec3} yields
an expression for the i-prob to observe detection events
which is identical to the probability distribution obtained from
the quantum theoretical description in terms of the singlet state
of two spin-1/2 particles~\cite{BOHM51,BALL03}.

\subsection{Experiment}\label{EPRBa}

\begin{figure}
\includegraphics[width=\hsize]{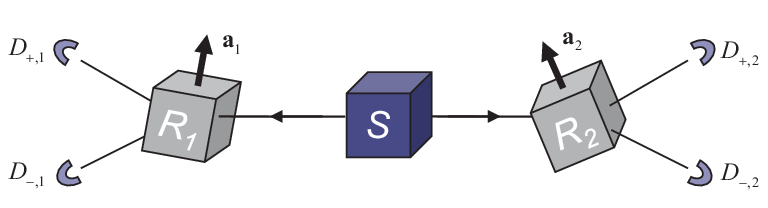}
\caption{(Color online)
Diagram of the EPRB thought experiment.
The source $S$, activated at times labeled by $n=1,2,\ldots,N$,
sends a signal to the router $R_1$ and another signal
to the router $R_2$.
Depending on the orientations of the routers,
represented by unit vectors $\mathbf{a_1}$ and $\mathbf{a_2}$,
the signal going to the left (right) is detected with 100\% certainty by either
$D_{+,1}$ or $D_{-,1}$ ($D_{+,2}$ or $D_{-,2}$).
}
\label{EPRBthought}
\end{figure}

The layout of the EPRB thought experiment is shown in Fig.~\ref{EPRBthought}.
In contrast to the conventional quantum theoretical description~\cite{BOHM51,BALL03},
we keep the number of assumptions about the experiment itself to a minimum.
Specifically, we assume that
\begin{enumerate}[a.]
\item
Each time the source $S$ is activated, it sends a signal to the left and another one to the right.
For the present purpose, it is not necessary to make any assumption about the nature of
or the correlation between these two signals.
\item
The observation station $i=1,2$ contains a ``router'' $R_i$ which sends the
signal to either detector $D_{+,i}$ or detector $D_{-,i}$.
The decision to send the signal to either $D_{+,i}$ or $D_{-,i}$
depends on the directions $\mathbf{a}_i$ of the router $R_i$,
$\mathbf{a}_i$ being a three-dimensional unit vector.
The orientations of the routers are relative to the fixed laboratory frame of reference.
\item
The detectors register the signal and operate with 100\% efficiency,
that is, if $n=1,2,\ldots,N$ labels the time at which the source is activated,
the firing of the detectors produces
a pair of integers $\{x_n,y_n\}$ where
$x_n=\pm1$ ($y_n=\pm1$) represents the firing of $D_{\pm,1}$ ($D_{\pm,2}$).
\end{enumerate}

The result of a run of the experiment for fixed $\mathbf{a}_1$ and $\mathbf{a}_2$ is a data set of pairs
\begin{eqnarray}
\Upsilon&=&\{x_n,y_n|x_n=\pm1;y_n=\pm1;n=1,\ldots,N \}
,
\label{prop0}
\end{eqnarray}
where $N$ is the total number of signal pairs emitted by the source.
From the data set Eq.~(\ref{prop0}), we compute the averages
\begin{eqnarray}
\langle x \rangle &=& \frac{1}{N}\sum_{i=1}^N x_i
\quad,\quad
\langle y \rangle = \frac{1}{N}\sum_{i=1}^N y_i
,
\label{prop0a}
\end{eqnarray}
the correlation and coincidences
\begin{eqnarray}
\langle xy \rangle &=& \frac{1}{N}\sum_{i=1}^N x_iy_i
\quad,\quad
n_{xy} = \sum_{i=1}^N \delta_{x,x_i}\delta_{y,y_i}
,
\label{prop0b}
\label{prop0c}
\\ \nonumber
\end{eqnarray}
which represents the number of events of the type $\{x,y\}$.
The assumptions (a--c) and Eqs.~(\ref{prop0a}) -- (\ref{prop0c})
represent our perception about the experiment and
specify the data analysis procedure, respectively.

\subsection{Inference-probability of the data produced by the experiment}\label{EPRBb}
The next step is to formalize the general features of the
possible outcomes of the experiment.

\begin{enumerate}[1.]
\item
The i-prob to observe a pair $\{x,y\}$
is denoted by $P(x,y|\mathbf{a}_1,\mathbf{a}_2,Z)$
where $Z$ represents all the conditions
under which the experiment is performed, with exception of
the directions $\mathbf{a}_1$ and $\mathbf{a}_2$ of the routers $R_1$ and $R_2$, respectively.
It is assumed that the conditions represented by $Z$ are fixed and identical
for all experiments.

It is not difficult to see that any real-valued function $f(x,y)$
of two dichotomic variables $x,y=\pm1$ can be written as
\begin{eqnarray}
f(x,y)&=&\frac{(1-x)(1-y)f(-1,-1)+(1+x)(1-y)f(+1,-1)}{4}
\nonumber \\&&+
\frac{(1-x)(1+y)f(-1,+1)+(1+x)(1+y)f(+1,+1)}{4}
\nonumber \\
&=&\frac{f(-1,-1)+f(+1,-1)+f(-1,+1)+f(+1,+1)}{4}
\nonumber \\
&&+x\frac{-f(-1,-1)+f(+1,-1)-f(-1,+1)+f(+1,+1)}{4}
\nonumber \\&&
+y\frac{-f(-1,-1)-f(+1,-1)+f(-1,+1)+f(+1,+1)}{4}
\nonumber \\&&
+xy\frac{f(-1,-1)-f(+1,-1)-f(-1,+1)+f(+1,+1)}{4}.
\nonumber \\
\label{prop1z}
\end{eqnarray}
From this general identity, it immediately follows that $P(x,y|\mathbf{a}_1,\mathbf{a}_2,Z)$ can be written as
\begin{eqnarray}
P(x,y|\mathbf{a}_1,\mathbf{a}_2,Z)=\frac{E_0(\mathbf{a}_1,\mathbf{a}_2,Z)+xE_1(\mathbf{a}_1,\mathbf{a}_2,Z) 
  +
  yE_2(\mathbf{a}_1,\mathbf{a}_2,Z)+xyE_{12}(\mathbf{a}_1,\mathbf{a}_2,Z)}{4}
,
\label{prop1}
\end{eqnarray}
where
\begin{eqnarray}
E_0(\mathbf{a}_1,\mathbf{a}_2,Z)&=&\sum_{x,y=\pm1}P(x,y|\mathbf{a}_1,\mathbf{a}_2,Z)=1,
\nonumber \\
E_1(\mathbf{a}_1,\mathbf{a}_2,Z)&=&\sum_{x,y=\pm1}xP(x,y|\mathbf{a}_1,\mathbf{a}_2,Z),
\nonumber \\
E_2(\mathbf{a}_1,\mathbf{a}_2,Z)&=&\sum_{x,y=\pm1}yP(x,y|\mathbf{a}_1,\mathbf{a}_2,Z),
\nonumber \\
E_{12}(\mathbf{a}_1,\mathbf{a}_2,Z)&=&\sum_{x,y=\pm1}xyP(x,y|\mathbf{a}_1,\mathbf{a}_2,Z).
\label{prop1a}
\end{eqnarray}
Furthermore, from Eq.~(\ref{prop1a}) and rule (a) (see section~\ref{sec2}), it follows directly that
$|E_1(\mathbf{a}_1,\mathbf{a}_2,Z)|\le 1$,
$|E_2(\mathbf{a}_1,\mathbf{a}_2,Z)|\le 1$,
and
$|E_{12}(\mathbf{a}_1,\mathbf{a}_2,Z)|\le 1$.
\item
For simplicity,
it is assumed that there is no relation between the actual values
of the pairs $\{x_n,y_n\}$ and $\{x_{n'},y_{n'}\}$ if $n\not=n'$.
In other words, as far as we know, each repetition
of the experiment represents an identical event of which
the outcome is logically independent of any other such event.
In probability theory, events with these properties are called Bernoulli trials, a concept that
is central to many results in probability theory~\cite{TRIB69,GRIM95,JAYN03}.
Invoking the product rule, the logical consequence of this
assumption is that
\begin{eqnarray}
P(x_1,y_1,\ldots,x_N,y_N|\mathbf{a}_1,\mathbf{a}_2,Z)
&=&
P(x_1,y_1|x_2,y_2,\ldots,x_N,y_N,\mathbf{a}_1,\mathbf{a}_2,Z)
P(x_2,y_2,\ldots,x_N,y_N|\mathbf{a}_1,\mathbf{a}_2,Z)
\nonumber \\
&=&
P(x_1,y_1|\mathbf{a}_1,\mathbf{a}_2,Z) P(x_2,y_2,\ldots,x_N,y_N|\mathbf{a}_1,\mathbf{a}_2,Z)
\nonumber \\
&=&
P(x_1,y_1|\mathbf{a}_1,\mathbf{a}_2,Z)
P(x_2,y_2|x_3,y_3,\ldots,x_N,y_N,\mathbf{a}_1,\mathbf{a}_2,Z)
\nonumber \\
&&\times
P(x_3,y_3,\ldots,x_N,y_N|\mathbf{a}_1,\mathbf{a}_2,Z)
\nonumber \\
&=&
P(x_1,y_1|\mathbf{a}_1,\mathbf{a}_2,Z)
P(x_2,y_2|\mathbf{a}_1,\mathbf{a}_2,Z)
P(x_3,y_3,\ldots,x_N,y_N|\mathbf{a}_1,\mathbf{a}_2,Z)
\nonumber \\
&=&\ldots
\nonumber \\
&=&
\prod_{i=1}^{N}P(x_i,y_i|\mathbf{a}_1,\mathbf{a}_2,Z)
,
\label{prop2}
\end{eqnarray}
meaning that the i-prob
$P(x_1,y_1,\ldots,x_N,y_N|\mathbf{a}_1,\mathbf{a}_2,Z)$
to observe the compound event $\left\{\{x_1,y_1\},\ldots,\{x_N,y_N\}\right\}$
is completely determined by the i-prob $P(x,y|\mathbf{a}_1,\mathbf{a}_2,Z)$ to observe the pair $\{x,y\}$.
\item
It is assumed that the i-prob $P(x,y|\mathbf{a}_1,\mathbf{a}_2,Z)$
to observe a pair $\{x,y\}$ does not change if we apply
the same rotation to both routers $R_1$ and $R_2$.
Expressing this invariance with respect to rotations
of the coordinate system (Euclidean space and Cartesian coordinates are used throughout this paper)
in terms of i-probs requires that
$P(x,y|\mathbf{a}_1,\mathbf{a}_2,Z)=P(x,y|{\cal R}\mathbf{a}_1,{\cal R}\mathbf{a}_2,Z)$
where ${\cal R}$ denotes an arbitrary rotation in three-dimensional space which is
applied to both routers $R_1$ and $R_2$.
As a function of the vectors $\mathbf{a}_1$ and $\mathbf{a}_2$,
the functional equation
$P(x,y|\mathbf{a}_1,\mathbf{a}_2,Z)=P(x,y|{\cal R}\mathbf{a}_1,{\cal R}\mathbf{a}_2,Z)$
can only be satisfied for all $\mathbf{a}_1$, $\mathbf{a}_2$ and
rotations ${\cal R}$ if $P(x,y|\mathbf{a}_1,\mathbf{a}_2,Z)$
is a function of the inner product $\mathbf{a}_1\cdot\mathbf{a}_2$ only.
Therefore, we must have
\begin{equation}
P(x,y|\mathbf{a}_1,\mathbf{a}_2,Z)=P(x,y|\mathbf{a}_1\cdot\mathbf{a}_2,Z)=P(x,y|\theta,Z)
,
\label{prop3}
\end{equation}
where $\theta=\arccos(\mathbf{a}_1\cdot\mathbf{a}_2)$ denotes the angle between the unit vectors
$\mathbf{a}_1$ and $\mathbf{a}_2$.
For any integer value of $K$, $\theta+2\pi K$
represents the same physical arrangement of the routers
$R_1$ and $R_2$.
\item
According to the basic rules of logical inference,
the i-prob to observe $x$, irrespective of the observed value of $y$ is given by
\begin{equation}
P(x|\mathbf{a}_1,\mathbf{a}_2,Z)=\sum_{y=\pm1}P(x,y|\mathbf{a}_1,\mathbf{a}_2,Z)
.
\label{prop3a}
\end{equation}
The assumption that observing $x=+1$ is as likely as
observing $x=-1$, independent of the observed value of $y$,
implies that we must have $P(x=+1|\mathbf{a}_1,\mathbf{a}_2,Z)=P(x=-1|\mathbf{a}_1,\mathbf{a}_2,Z)$ which,
in view of the fact that $P(x=+1|\mathbf{a}_1,\mathbf{a}_2,Z)+P(x=-1|\mathbf{a}_1,\mathbf{a}_2,Z)=1$
implies that $P(x=+1|\mathbf{a}_1,\mathbf{a}_2,Z)=P(x=-1|\mathbf{a}_1,\mathbf{a}_2,Z)=1/2$.
Applying the same reasoning to the assumption that, independent of the observed values of $x$,
observing $y=+1$ is as likely as observing $y=-1$
yields $P(y|\mathbf{a}_1,\mathbf{a}_2,Z)=P(x=+1,y|\mathbf{a}_1,\mathbf{a}_2,Z)+P(x=-1,y|\mathbf{a}_1,\mathbf{a}_2,Z)=1/2$.
Then, from Eq.~(\ref{prop1a}) it follows directly that
\begin{eqnarray}
E_1(\mathbf{a}_1,\mathbf{a}_2,Z)&=&E_2(\mathbf{a}_1,\mathbf{a}_2,Z)=0
.
\label{prop4}
\end{eqnarray}
\end{enumerate}

\noindent
Assumptions (3) and (4) formalize our expectations
about the symmetries of the experimental setup.
Note that we did not assign any prior i-prob nor that
at this stage, there is any reference to concepts such
as the singlet-state.
Although the symmetry properties which have been assumed
are reminiscent of those of the singlet state,
this is deceptive.
As we show later, without altering the assumptions
that are expressed in (3) and (4),
the logical-inference approach yields the
correlations for triplet states as well.
Using Eqs.~(\ref{prop1}),~(\ref{prop1a}),~(\ref{prop3}), and (\ref{prop4}) we find that
the i-prob to observe a pair $\{x,y\}$ simplifies to
\begin{equation}
P(x,y|\theta,Z)=\frac{1+xyE_{12}(\theta)}{4}
,
\label{prop5}
\end{equation}
where
$E_{12}(\theta)=E_{12}(\mathbf{a}_1,\mathbf{a}_2,Z)$ is a periodic function of $\theta$.

\subsection{Condition for reproducibility and robustness}\label{EPRBc}

Although the data set Eq.~(\ref{prop0}) changes from run to run,
we expect that the averages Eq.~(\ref{prop0a}), the correlation and the coincidences Eq.~(\ref{prop0c})
exhibit some kind of robustness, a smoothness with respect to small changes of $\theta$.
If this were not the case, these numbers would vary
erratically with  $\theta$. Most likely the results would be called ``irreproducible'',
and the experimental data would be disposed of because
repeating the run with a slightly different value of $\theta$
would often produce results that are very different from those of other runs.

Obviously, the important feature of robustness with respect to small variations
of the conditions under which the experiment is carried out
should be reflected in the expression
for the i-prob to observe data sets which yield reproducible
averages and correlations (with the usual statistical fluctuations).
Having exploited all elementary knowledge about the experiment
(see Sections~\ref{EPRBa} and \ref{EPRBb}),
the next step therefore is to determine the expression
for $P(x,y|\theta,Z)$ which is most insensitive to small changes in $\theta$.

Let us assume that for a fixed value of $\theta$,
an experimental run of $N$ events yields
$n_{xy}$ events of the type $\{x,y\}$
where $n_{++}+n_{-+}+n_{+-}+n_{--}=N$.
The number of different data sets yielding the same
values of $n_{++}$, $n_{-+}$, $n_{+-}$, and $n_{--}$
is $N!/(n_{++})!(n_{-+})!(n_{+-})!(n_{--})!$.
According to Eq.~(\ref{prop2}), the i-prob that events of the type $\{x,y\}$
occur $n_{xy}$ times is given by
$\prod_{x,y=\pm1} P(x,y|\theta,Z)^{n_{xy}}$.
Therefore, the i-prob to observe
the (compound) event $\{n_{++},n_{-+},n_{+-},n_{--}\}$
is given by
\begin{equation}
P(n_{++},n_{-+},n_{+-},n_{--}|\theta,N,Z)=
N!\prod_{x,y=\pm1} \frac{P(x,y|\theta,Z)^{n_{xy}}}{n_{xy}!}
.
\label{robu0}
\end{equation}

If the outcome of the experiment is indeed described by the i-prob
Eq.~(\ref{robu0}) and the experiment is supposed
to yield reproducible, robust results, small changes
of $\theta$ should not have a drastic effect on the outcome.
So let us ask ourselves how the i-prob would change if the experiment is carried
out with $\theta+\epsilon$ ($\epsilon$ small) instead of with $\theta$.

It is expedient to formulate this question as an hypothesis test.
Let $H_0$ and $H_1$ be the hypothesis that
the data $\{n_{++},n_{-+},n_{+-},n_{--}\}$ is observed if
the angle between the unit vectors $\mathbf{a}_1$ and $\mathbf{a}_2$
is $\theta$ and $\theta+\epsilon$, respectively.
The evidence $\mathrm{Ev}$ of hypothesis $H_1$, relative to hypothesis $H_0$,
is defined by~\cite{TRIB69,JAYN03}
\begin{equation}
\mathrm{Ev}=\ln\frac{P(n_{++},n_{-+},n_{+-},n_{--}|\theta+\epsilon,N,Z)}{P(n_{++},n_{-+},n_{+-},n_{--}|\theta,N,Z)}
,
\label{robu1}
\end{equation}
where the logarithm serves to facilitate the algebraic manipulations.
If $H_1$ is more (less) plausible than $H_0$ then $\mathrm{Ev}>0$ ($\mathrm{Ev}<0$).

The absolute value of the evidence, $|\mathrm{Ev}|$ is a measure for the robustness of the description
(the sign of $\mathrm{Ev}$ is arbitrary, hence irrelevant).
The problem of determining the most robust description of the experimental data may now be formulated
as follows: search for the i-prob's $P(n_{++},n_{-+},n_{+-},n_{--}|\theta,N,Z)$
which minimize $|\mathrm{Ev}|$ for all possible $\epsilon$ ($\epsilon$ small) and for all possible $\theta$.
The condition ``for all possible $\epsilon$ and  $\theta$'' renders the
minimization problem an instance of a robust optimization problem~\cite{WIKIROBUST}.

Obviously, this robust optimization problem has a trivial solution, namely
$P(n_{++},n_{-+},n_{+-},n_{--}|\theta,N,Z)$ independent of $\theta$.
For the case at hand, such $P(n_{++},n_{-+},n_{+-},n_{--}|\theta,N,Z)$'s
can only describe experiments for which $\{n_{++},n_{-+},n_{+-},n_{--}\}$
does not exhibit any dependence on $\theta$, the angle between
the vectors $\mathbf{a}_1$ and $\mathbf{a}_2$ which represent
the direction of the routers $R_1$ and $R_2$.
\begin{center}
\framebox{
\parbox[t]{0.9\hsize}{%
Experiments which produce results that do not change with
the conditions do not increase our knowledge about the relation
between the conditions and the observed data.
In this paper, we do not consider such (fairly useless) experiments
and consequently, we explicitly exclude solutions
for the i-probs that are constant with respect to the conditions.
}}
\end{center}

Assume that we have found a set of i-prob's $P_0(n_{++},n_{-+},n_{+-},n_{--}|\theta,N,Z)$ which are not all constant functions of $\theta$
and which minimize $|\mathrm{Ev}|$ for all $\epsilon$ ($\epsilon$ small).
Call this minimum $|\mathrm{Ev}|_0$.
Suppose that $|\mathrm{Ev}|_0$ itself depends on $\theta$, meaning that the robustness of the description varies with $\theta$.
Then, the found set of i-prob's is definitely not a solution of the robust optimization problem
because it does not satisfy the condition that the solution must hold for {\sl all} possible $\theta$.
Therefore, to solve the problem induced by the ``for all possible $\theta$'' clause,
we require that at the minimum, $|\mathrm{Ev}|$ is independent of $\theta$.

Summarizing: our concept of a robust experiment implies that the i-prob's which
describe such experiment can be found by minimizing $|\mathrm{Ev}|$,
subject to the constraints that
\begin{enumerate}[(C1)]
\item
$\epsilon$ is small but arbitrary.
\item
Not all i-prob's are independent of $\theta$.
\item
$|\mathrm{Ev}|$ is independent of $\theta$.
\end{enumerate}
Although we have used a particular example to introduce and illustrate the
concept of robustness, we will use the same notion throughout this paper.

\subsection{Robust solution}\label{EPRBc0}

Making use of Eq.~(\ref{robu0}), we find
\begin{equation}
\mathrm{Ev}=\sum_{x,y=\pm1}n_{xy}\ln\frac{P(x,y|\theta+\epsilon,Z)}{P(x,y|\theta,Z)}
.
\label{robu2}
\end{equation}
Writing Eq.~(\ref{robu2}) as a Taylor series in $\epsilon$ we have
\begin{eqnarray}
\mathrm{Ev}&=&
\sum_{x,y=\pm1}n_{xy}\left\{
\epsilon\frac{P'(x,y|\theta,Z)}{P(x,y|\theta,Z)}
-\frac{\epsilon^2}{2}\left(\frac{P'(x,y|\theta,Z)}{P(x,y|\theta,Z)}\right)^2
+\frac{\epsilon^2}{2}\frac{P''(x,y|\theta,Z)}{P(x,y|\theta,Z)}
\right\}
+{\cal O}(\epsilon^3)
,
\label{robu3}
\end{eqnarray}
where a prime denotes the derivative with respect to the variable $\theta$.

According to our notion of robustness,
the evidence Eq.~(\ref{robu3}) should change as little as possible
as $\epsilon$ varies.
The contribution of the term in $\epsilon$ can be made to vanish
by substituting $n_{xy}=\alpha P(x,y|\theta,Z)$.
From
\begin{eqnarray}
N&=&\sum_{x,y=\pm1}n_{xy}=\alpha \sum_{x,y=\pm1} P(x,y|\theta,Z)=\alpha
,
\label{robu4a}
\end{eqnarray}
it follows that $\alpha=N$.
Then we have
\begin{eqnarray}
\sum_{x,y=\pm1}n_{xy}\frac{P'(x,y|\theta,Z)}{P(x,y|\theta,Z)}
&=&
N\sum_{x,y=\pm1}P'(x,y|\theta,Z)
\nonumber \\&=&
N\frac{\partial}{\partial\theta}\sum_{x,y=\pm1}P(x,y|\theta,Z)
\nonumber \\&=&
N\frac{\partial}{\partial\theta} 1 = 0
.
\label{robu4}
\end{eqnarray}
Using the same reasoning, it follows that the third term
in Eq.~(\ref{robu3}) also vanishes
and we have

\begin{equation}
\mathrm{Ev}=
-\frac{N\epsilon^2}{2}\sum_{x,y=\pm1}
\frac{1}{P(x,y|\theta,Z)}
\left(\frac{\partial P(x,y|\theta,Z)}{\partial\theta}\right)^2
+{\cal O}(\epsilon^3)
.
\label{robu5}
\end{equation}
Although our choice $P(x,y|\theta,Z)=n_{xy}/N$
is motivated by the desire to eliminate contributions of order $\epsilon$,
it follows that our criterion of robustness
leads us to the intuitively obvious procedure which assigns to $P(x,y|\theta,Z)$
the value of the observed frequencies of occurrences $n_{xy}/N$.
As shown in \ref{APP1}, for large $N$, the same procedure also follows
from searching for the $P(x,y|\theta,Z)$'s which maximize
the i-prob to observe $\{n_{++},n_{-+},n_{+-},n_{--}\}$.

Omitting terms of ${\cal O}(\epsilon^3)$, minimizing
$|\mathrm{Ev}|$ while taking into account the
constraints (C2) and (C3) (see Section~\ref{EPRBc})
amounts to finding the i-prob's
$P(x,y|\theta,Z)$ which minimize
\begin{equation}
I_{F}= \sum_{x,y=\pm1}
\frac{1}{P(x,y|\theta,Z)}
\left(\frac{\partial P(x,y|\theta,Z)}{\partial\theta}\right)^2
,
\label{robu6}
\end{equation}
subject to the constraint that $\partial P(x,y|\theta,Z)/\partial\theta\not=0$
for some pairs $(x,y)$.
The r.h.s. of Eq.~(\ref{robu6}) is the Fisher information
for the problem at hand and because of constraint (C3), does not depend on $\theta$.

Using Eq.~(\ref{prop5}), we can rewrite Eq.~(\ref{robu6}) as
\begin{equation}
I_F=\frac{1}{1-E_{12}^2(\theta)}
\left(\frac{\partial E_{12}(\theta)}{\partial \theta}\right)^2
,
\label{robu6a}
\end{equation}
which is readily integrated to yield

\begin{equation}
E_{12}(\theta)=\cos\left(\theta\sqrt{I_F}+\phi\right)
.
\label{robu7}
\end{equation}
where $\phi$ is an integration constant.

As $E_{12}(\theta)$ is a periodic function of $\theta$
we must have $\sqrt{I_F}=K$ where $K$ is an integer
and hence
\begin{equation}
E_{12}(\theta)=\cos\left(K\theta+\phi\right)
.
\label{robu8}
\end{equation}

Because of constraint (C2)
we exclude the case $K=I_F=0$ from further consideration
because it describes an experiment in which
the frequency distribution of the observed data
does not depend on $\theta$.
Therefore, the physically relevant,
nontrivial solution with minimum Fisher information
corresponds to $K=1$.
Furthermore, as $E_{12}(\theta)$ is a function of
$\mathbf{a}_1\cdot\mathbf{a}_2=\cos\theta$ only,
we must have $\phi=0,\pi$, reflecting an ambiguity
in the definition of the direction
of $R_1$ relative to the direction of $R_2$.

Choosing the solution with $\phi=\pi$,
the two-particle correlation reads
\begin{eqnarray}
E_{12}(\mathbf{a}_1,\mathbf{a}_2,Z)
&=&-\cos\theta=-\mathbf{a}_1\cdot\mathbf{a}_2,
\label{robu9}
\end{eqnarray}
in agreement with the expression for the correlation of two $S=1/2$
particles in the singlet state~\cite{BOHM51,BALL03}.

\subsection{Discussion}\label{EPRBd}
We have shown that the application of our criterion of robust, reproducible experiments
to the EPRB thought experiment depicted in Fig.~\ref{EPRBthought}
amounts to minimizing the Fisher information Eq.~(\ref{robu7}) for this specific problem.
The result of this calculation is the correlation Eq.~(\ref{robu9})
which is characteristic for the singlet state.
Needless to say, our derivation did not use any concepts of quantum theory.
Only plain, rational reasoning strictly complying with the rules
of logical inference and some elementary facts about the experiment were used.

It is most remarkable that the equations of quantum theory
for a system in the singlet state appear by simply
requiring that (i) everything which is known about the source
is uncertain, except that it emits two signals, (ii) the routers
$R_1$ and $R_2$ transform the received signal
into two-valued signals, and that (iii) the i-prob describing the frequencies of the observed events depends
on the relative orientation of the routers only,
see Eq.~(\ref{prop3}).
Apparently, the latter requirement suffices to recover the
salient feature of the singlet state of two spin-1/2 particles, namely that
the state vector
$|\psi\rangle=\left(|\uparrow\downarrow\rangle-|\downarrow\uparrow\rangle\right)/\sqrt{2}$
is invariant for rotations, implying that its physical properties
do not depend on the direction chosen to define ``up'' or ``down''~\cite{BOHM51,BALL03}.
Realizing conditions (i) and (iii)
in a real EPRB experiment is not a trivial matter~\cite{WEIH00,SHIH11,VIST12}.

The correlations that are characteristic for other entangled states
for which Eq.~(\ref{prop4}) holds are obtained
by making different assumptions about the properties of the routers.
As an example, assume that the output of router $R_1$
is determined by  $(-a^x,a^y,-a^z)$ instead of by $(a^x,a^y,a^z)$.
Repeating the derivation that leads to Eq.~(\ref{robu9})
yields $E_{12}(\mathbf{a}_1,\mathbf{a}_2,Z)= + a_1^x a_2^x - a_1^y a_2^y + a_1^z a_2^z$,
which, in quantum theory, is the correlation
of two $S=1/2$ particles described by
the state vector $|\psi\rangle=\left(|\uparrow\uparrow\rangle+|\downarrow\downarrow\rangle\right)/\sqrt{2}$,
the triplet state with the z-projection of total magnetization zero.

\section{Stern-Gerlach experiment}\label{SG}

\begin{figure}
\includegraphics[width=\hsize]{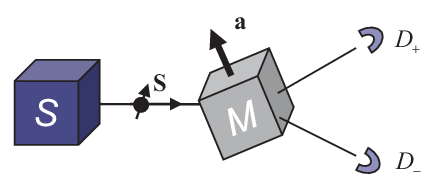}
\caption{(Color online)
Diagram of the Stern-Gerlach experiment.
The source $S$, activated at times labeled by $n=1,2,\ldots,N$,
sends a particle carrying a magnetic moment $\mathbf{S}$ to the magnet $M$ with
its magnetization in the direction $\mathbf{a}$.
Depending on the relative directions of $\mathbf{a}$ and $\mathbf{S}$,
the particle is detected with 100\% certainty by either
$D_{+}$ or $D_{-}$.
}
\label{SGthought}
\end{figure}

The expression Eq.~(\ref{robu9})
for the correlation of the data produced by an EPRB experiment
has been obtained without making specific assumptions
about the nature of the signals.
In this section, we add some extra assumptions and we show how the same reasoning of Section~\ref{EPRB}
leads to the expression of a simple quantum mechanical model of the Stern-Gerlach experiment, see Fig.~\ref{SGthought}.
In order to avoid repetition, in the following
we leave out arguments/derivations/discussions which, with minor changes
have been given earlier.

\subsection{Experiment}\label{SGa}

We start by listing the assumptions about the nature
of the signal and the action of the magnet on the signal.
Specifically, we assume that
\begin{enumerate}[a.]
\item
The signal emitted by the source takes the form of a particle
which carries a magnetic moment represented by a unit vector $\mathbf{S}$.
The magnetic moment interacts with the magnetic field generated by the magnet $M$.
This field is a function of the direction $\mathbf{a}$ of the magnet only.
The direction of the magnetic moment and magnet are relative to the fixed laboratory frame of reference.
\item
As the particle passes through the magnetic field, it is directed towards either $D_{+}$ or $D_{-}$.
The mechanism which causes this to happen is assumed to depend on $\mathbf{a}\cdot \mathbf{S}=\cos\theta$ only.
In other words, the distribution of the number of particles detected by $D_{+}$ or $D_{-}$
does not change (within the usual statistical fluctuations) if both the magnetic moment of the particles
and the direction of the magnetic field are rotated by the same amount.
Obviously, this is just expressing the assumption that space is isotropic.
\item
The detectors count the particles with 100\% efficiency,
that is, if $n=1,2,\ldots,N$ labels the time at which the source is activated,
the firing of the detectors produces a data set of integers $\{x_n|x_n=\pm1;n=1,\ldots,N\}$ where
$x_n=\pm1$ represents the firing of $D_{\pm}$.
\end{enumerate}

\subsection{Inference-probability of the data produced by the experiment}\label{SGb}

In complete analogy with Section~\ref{EPRBb}, we have
\begin{enumerate}[1.]
\item{
The i-prob to observe an event $x=\pm1$
is denoted by $P(x|\mathbf{a}\cdot\mathbf{S},Z)$
where $Z$ represents all the conditions
under which the experiment is performed, with exception of
the directions $\mathbf{a}$ of the magnet and $\mathbf{S}$ of the magnetic moment of the particle.
It is assumed that the conditions represented by $Z$ are fixed and identical
for all experiments.
It is expedient to write $P(x|\mathbf{a}\cdot\mathbf{S},Z)$ as
\begin{equation}
P(x|\mathbf{a}\cdot\mathbf{S},Z)=P(x|\theta,Z)=\frac{1+xE(\theta)}{2}
,
\label{sg1}
\end{equation}
where
\begin{eqnarray}
E(\theta)&=&E(\mathbf{a}\cdot\mathbf{S},Z)=\sum_{x=\pm1}xP(x|\theta,Z)
.
\label{sg1a}
\end{eqnarray}
}
\item{
For simplicity, it is assumed that there is no relation between the actual values
of $x_n$ and $x_{n'}$ if $n\not=n'$.
In other words, as far as we know, each repetition
of the experiment represents an identical event of which
the outcome is logically independent of any other such event.
Repeated application of the product rule yields
\begin{eqnarray}
P(x_1,\ldots,x_N|\mathbf{a}\cdot\mathbf{S},Z)
&=&
\prod_{i=1}^{N}P(x_i|\theta,Z)
,
\label{sg2}
\end{eqnarray}
meaning that the i-prob  $P(x|\mathbf{a}\cdot\mathbf{S},Z)$ to observe the event $\{x_1,\ldots,x_N\}$
is uniquely determined by the i-prob to observe the event $x$.
}
\end{enumerate}

\subsection{Condition for reproducibility and robustness}\label{SGc}

Enforcing the condition of reproducibility,
exactly the same reasoning that leads to Eq.~(\ref{robu5})
now yields
\begin{equation}
\mathrm{Ev}=
-\frac{N\epsilon^2}{2}\sum_{x=\pm1}
\frac{1}{P(x|\theta,Z)}
\left(\frac{\partial P(x|\theta,Z)}{\partial\theta}\right)^2
+{\cal O}(\epsilon^3)
,
\label{sg3}
\end{equation}
from which it follows that in order to reduce
the variation of Eq.~(\ref{sg3}) as a function of $\epsilon$
as much as possible, we should minimize the Fisher information
\begin{eqnarray}
I_{F}&=&\sum_{x=\pm1}
\frac{1}{P(x|\theta,Z)}
\left(\frac{\partial P(x|\theta,Z)}{\partial\theta}\right)^2
\nonumber \\
&=&
\frac{1}{1-E^2(\theta)}
\left(\frac{\partial E(\theta)}{\partial \theta}\right)^2
,
\label{sg4}
\end{eqnarray}
subject to the constraint $\partial  P(x|\theta,Z)/\partial\theta\not=0$.
The method of solution is identical to the one
employed in Section~\ref{EPRBc0}.
Using the fact that $E(\theta)$ is a function
of $\mathbf{a}\cdot\mathbf{S}=\cos\theta$
only, we find that there are two
solutions, namely $E(\theta)=\pm \cos\theta$.
Therefore, we have
\begin{eqnarray}
P(x|\mathbf{a}\cdot\mathbf{S},Z)=
P(x|\theta,Z)=\frac{1\pm x\mathbf{a}\cdot\mathbf{S}}{2}
,
\label{sg5}
\end{eqnarray}
in agreement with the expressions of the quantum theoretical
expression for the probability to deflect the particle
in one of the two distinct directions labeled by $x=\pm1$~\cite{BALL03}.
The $\pm$ sign in Eq.~(\ref{sg5}) reflects the fact that
the mapping between $x=\pm1$ and the two different directions
is only determined up to a sign.

\subsection{Discussion}\label{SGd}

In quantum theory, Eq.~(\ref{sg5}) is in essence just the postulate (Born's rule)
that the probability to observe the particle with spin up (corresponding to say $x=+1$)
is given by the square of the absolute value of the amplitude of the wavefunction
projected onto the spin-up state~\cite{BALL03}.
Obviously, the variability of the conditions under which
an experiment is carried out  is not included in the quantum theoretical description.
In contrast, in the logical inference approach, Eq.~(\ref{sg5}) is not postulated
but follows from the assumption that the (thought) experiment that is being
performed yields the most reproducible results,
revealing the conditions for an experiment to produce data which
is described by quantum theory.

\section{Particle in a potential: Schr\"odinger equation}\label{SE}

Sections~\ref{EPRB} and ~\ref{SG} showed that, with a minimum of input about the nature of an experiment,
simply demanding that the recorded data sets of events yield reproducible results
for the i-probs, leads to expressions that are known from
the quantum theoretical treatment of the experiment.
In essence, these results derive from the following ideas:
\begin{enumerate}[(i)]
\item
The i-probs for events to occur obey the rules of the algebra of logical inference.
\item
The i-prob to observe an event (labeled by $\{x,y\}$ or $x$) depends explicitly on a variable condition
(represented by the variable $\theta$).
\item
Maximizing the robustness of the i-prob to observe the data
with respect to small variations of the condition
yields the functional dependence of the i-prob on this condition.
\end{enumerate}
This section shows that extending this approach to a particle in a potential is straightforward.
The key points are to formulate precisely what it means to perform a robust, reproducible experiment
and to feed in knowledge about the Newtonian dynamics of the particle.
We consider the time-independent case and
to keep the notation simple, we only treat the case of a particle on a line.
The extension to 2- or 3-dimensional space and the time-dependent case is
given in Section~\ref{TDSE}.

\subsection{Experiment}\label{SEa}

We consider the following experiment.
A particle is located on a line segment $[-L,L]$,
relative to a fixed reference frame.
Its unknown position is denoted by $\theta$.
We have a source that emits a signal
which always solicits a response of the particle.
We cover another line segment $[-L,L]$
with $2M+1$ detectors of width $\Delta$, where $M\Delta=L$.
The signal that arrives at detector $j$ with $-M\le j\le M$ is assumed to be particle-like, that is
for each signal emitted by the source, only one of the $2M+1$ detectors actually fires.
Each detector operates with 100\% efficiency,
meaning that it fires whenever a particle-like signal arrives.

The result of a run of the experiment is a data set of detector clicks
\begin{eqnarray}
\Upsilon&=&\{j_n|-M\le j_n \le M;\; n=1,\ldots,N \}
.
\label{SE0}
\end{eqnarray}
Denoting the total count of detector $j$ by $0\le k_j\le N$,
the experiment produces the data set
\begin{eqnarray}
{\cal D}&=&\left\{k_{-M},\ldots,k_{M}|N=k_{-M}+\ldots+k_{M} \right\}
.
\label{SE1}
\end{eqnarray}

\subsection{Inference-probability of the data produced by the experiment}\label{SEb}

A priori, the relation between the unknown location $\theta$ of the particle
and the location $j$ of the detector which fires is unknown.
Therefore, to describe this relation,
we introduce the i-prob $P(j|\theta,Z)$ that the particle at unknown location $\theta$
activates the detector located at the position $-M\le j \le M$.
As before, the conditions represented by $Z$ are fixed and identical for all experiments.
As in Sections~\ref{EPRB} and ~\ref{SG}, the key question is what the requirement of
reproducibility tells us about the i-prob $P(j|\theta,Z)$ as a function of $\theta$.
Note that unlike in the case of parameter estimation,
in the case at hand both $P(j|\theta,Z)$ and the parameter $\theta$ are unknown.

The following assumptions are essentially the same as those of  Sections~\ref{EPRBb} and ~\ref{SGb}
and are listed here for completeness.
\begin{enumerate}[1.]
\item
For fixed position $\theta$, the i-prob to observe the data
is given by
\begin{eqnarray}
P({\cal D}|\theta,N,Z)
&=& P(k_{-M},\ldots,k_{M}|\theta,N,Z)
.
\label{SE1a}
\end{eqnarray}
It is assumed that there is no relation between the actual values
of $j_n$ and $j_{n'}$ if $n\not=n'$.
In other words, each repetition of the experiment represents an identical event of which
the outcome is logically independent of any other such event.
As mentioned before, events with these properties are called Bernoulli trials, a concept which
is central to many results in probability theory~\cite{TRIB69,GRIM95,JAYN03}.
By the standard combinatorial argument,
the number of possible ways $N_{\cal D}$ to generate the data set ${\cal D}$ is given by
\begin{eqnarray}
N_{\cal D}
&=& \frac{N!}{k_{-M}!\ldots k_{M}!}
.
\label{SE1b}
\end{eqnarray}
The logical consequence of the Bernoulli-trial assumption is then that
\begin{eqnarray}
P({\cal D}|\theta,N,Z)
&=&P(k_{-M}|k_{-M+1},\ldots,k_{M},\theta,N,Z)
\nonumber \\
&&\times P(k_{-M+1},\ldots,k_{M}|\theta,N,Z)
\nonumber \\
&=&P(k_{-M}|\theta,N,Z)P(k_{-M+1},\ldots,k_{M}|\theta,N,Z)
\nonumber \\
&=&\ldots
\nonumber \\
&=& N_{\cal D} P(-M|\theta,N,Z)^{k_{-M}}\ldots P(M|\theta,N,Z)^{k_{M}}
\nonumber \\
&=& N!\prod_{j=-M}^M \frac{P(j|\theta,Z)^{k_j}}{k_j!}
.
\label{SE2}
\end{eqnarray}
\item
In physics we often assume that space is homogeneous,
implying that it does not matter where in space we perform the experiment.
For the model at hand, this means that a translation
of the unknown position $\theta$ and the array of detectors by
the same distance should not affect our inferences based on the data.
In other words, the i-prob has the property
\begin{equation}
P(j|\theta,Z)=P(j+\zeta|\theta+\zeta,Z)
,
\label{SE4}
\end{equation}
where $\zeta$ is an arbitrary real number.
\end{enumerate}

\subsection{Condition for reproducibility and robustness}\label{SEc}

Comparing Eq.~(\ref{SE2}) and Eq.~(\ref{robu0}),
it is not a surprise that
by simply repeating all the steps that
lead to Eq.~(\ref{robu5}),
the condition for reproducibility applied
to Eq.~(\ref{SE2}) yields the evidence
\begin{equation}
\mathrm{Ev}=
-\frac{N\epsilon^2}{2}\sum_{j=-M}^M
\frac{1}{P(j|\theta,Z)}
\left(\frac{\partial P(j|\theta,Z)}{\partial\theta}\right)^2
+{\cal O}(\epsilon^3)
.
\label{SE5}
\end{equation}
At this point, to make contact with the
Schr\"odinger equation which is formulated in continuum space,
it is necessary to replace Eq.~(\ref{SE5}) by its continuum limit
\begin{equation}
\mathrm{Ev}=
-\frac{N\epsilon^2}{2}\int_{-\infty}^\infty dx\;
\frac{1}{P(x|\theta,Z)}
\left(\frac{\partial P(x|\theta,Z)}{\partial\theta}\right)^2
+{\cal O}(\epsilon^3)
,
\label{SE6a}
\end{equation}
where we have assumed that the width of the detectors approaches zero ($\Delta\rightarrow0$)
and the length of the line segment approaches infinity ($L\rightarrow\infty$).
Making use of translational invariance (see Eq.~(\ref{SE4})) we have
\begin{eqnarray}
\frac{\partial P(x|\theta,Z)}{\partial\theta}
&=&\lim_{\delta\rightarrow 0}\frac{P(x|\theta+\delta,Z)-P(x|\theta,Z)}{\delta}
\nonumber \\
&=&\lim_{\delta\rightarrow 0}\frac{P(x-\delta|\theta,Z)-P(x|\theta,Z)}{\delta}
\nonumber \\
&=&-\frac{\partial P(x|\theta,Z)}{\partial x}
.
\label{SE6aa}
\end{eqnarray}
Hence, we may replace the partial derivative with respect to $\theta$
by the partial derivative with respect to $x$, yielding
\begin{equation}
\mathrm{Ev}=
-\frac{N\epsilon^2}{2}I_F
+{\cal O}(\epsilon^3)
,
\label{SE6b}
\end{equation}
where
\begin{equation}
I_F=\int_{-\infty}^\infty dx\;
\frac{1}{P(x|\theta,Z)}
\left(\frac{\partial P(x|\theta,Z)}{\partial x}\right)^2
,
\label{SE7}
\end{equation}
denotes the Fisher information of the experiment considered in this section.
Obviously, minimizing Eq.~(\ref{SE7}) as we did for the EPRB and Stern-Gerlach problem
cannot yield a solution which incorporates the fact that the particle moves
in a potential simply because this knowledge is not yet built into the minimization problem.

According to classical mechanics
the orbit in phase space of a particle is given by the solution of the
time-independent Hamilton-Jacobi equation (HJE)
\begin{equation}
\frac{1}{2m}\left(\frac{\partial S(\theta)}{\partial \theta}\right)^2+V(\theta)-E=0
,
\label{SE7a}
\end{equation}
where  $m$, $S(\theta)$, $V(\theta)$ and $E$
denote the mass of the particle,
the action (Hamilton's principal function),
the potential, and the energy, respectively.
Note that $\theta$ represents the position of the particle
which, in classical mechanics, is assumed to be known with certainty.
The HJE describes
experiments for which there is no uncertainty about each individual event (category 1).

If there is uncertainty about the position $\theta$
but not about the individual event (category 2 experiment),
this uncertainty may be captured by assuming
that the i-prob $P(x|\theta,Z)$,
has a particular functional dependence,
e.g. $P(x|\theta,Z)=\exp[-(x-\theta)^2/2\sigma^2]/\sqrt{2\pi\sigma^2}$.
Given that the equation which determines the action $S(x)$ should
reduce to Eq.~(\ref{SE7a}) in the limit that
$P(x|\theta,Z)\rightarrow \delta(x-\theta)$,
the simplest equation reads
\begin{equation}
\int_{-\infty}^\infty dx\;
\left[
\left(\frac{\partial S(x)}{\partial x}\right)^2
+2m[V(x)-E]
\right]P(x|\theta,Z)=0
.
\label{SE7b}
\end{equation}
In words, Eq.~(\ref{SE7b}) tells us that the inference drawn from
the distribution of detector clicks as function
of their location on the line,
is that, {\it on average}, these locations satisfy the time-independent HJE.

Finally, if there is uncertainty
about the individual events as well as about the conditions (category 3 experiment),
the i-prob $P(x|\theta,Z)$ is unknown but can be determined
by requiring that the frequency distributions of the observed events are robust (category 3a experiment).
It is important to note that in this case, there is no assumption
about the {\it unknown} position $\theta$ of the particle.

Inspired by Schr\"odinger's original  derivation~\cite{SCHR26a}
of his equation (see Section~\ref{SEz}),
we minimize the Fisher information Eq.~(\ref{SE7}) with the
constraint that the time-independent HJE only holds
on average.
Specifically, the functional to be minimized
under the constraint that $\partial P(x|\theta,Z)/\partial\theta=-\partial P(x|\theta,Z)/\partial x \not=0$
reads
%
\begin{eqnarray}
F(\theta)=
\int_{-\infty}^\infty dx\;
\left\{
\frac{1}{P(x|\theta,Z)}
\left(\frac{\partial P(x|\theta,Z)}{\partial x}\right)^2
+\lambda
\left[
\left(\frac{\partial S(x)}{\partial x}\right)^2
+2m[V(x)-E]
\right]P(x|\theta,Z)
\right\}
,
\nonumber \\
\label{SE8}
\end{eqnarray}
where $\lambda$ is a Lagrange multiplier.
It is important to note that without changing the minimization
problem, we may substitute
$P(x|\theta,Z) \rightarrow \alpha P(x|\theta,Z)$
where $\alpha$ is any nonzero real number.
Therefore, any solution for $P(x|\theta,Z)$ obtained by minimizing
Eq.~(\ref{SE8}) can be normalized by
$P(x|\theta,Z) \rightarrow P(x|\theta,Z)/\int_{-\infty}^\infty dx\;P(x|\theta,Z)$.
Hence, there is no need to introduce a Lagrange multiplier to impose the
normalization condition on $P(x|\theta,Z)$.

It is easy to show, directly from Eq.~(\ref{SE8}), that at an extremum
(with respect to variations in $P(x|\theta,Z)$ and $S(x)$, not to $\theta$)
the derivative of $F(\theta)$ with respect to $\theta$ is zero, that is
\begin{equation}
\left.\frac{\partial F(\theta)}{\partial \theta}\right|
_{\mathrm{Extremum}\;\mathrm{of}\;F(\theta)} =0
,
\label{SE12}
\end{equation}
hence the solutions of the variational problem
comply with the constraint (C3) (see Section~\ref{EPRBc}).

We do not know of any direct analytical method to solve
the nonlinear minimization problem Eq.~(\ref{SE8}).
However, from Madelung's hydrodynamic-like formulation~\cite{MADE27} or
Bohm's interpretation~\cite{BOHM52} of quantum theory
it follows that the extrema (and therefore also the minima)
of Eq.~(\ref{SE8}) can be found by solving the time-independent Schr\"odinger equation.

With a minimum of algebra this can be shown as follows.
We start from the functional
\begin{equation}
Q(\theta)=
\int_{-\infty}^\infty dx\;
\left\{
4
\frac{\partial \psi^\ast(x|\theta,Z)}{\partial x}
\frac{\partial \psi(x|\theta,Z)}{\partial x}
+2m\lambda[V(x)-E]\psi^\ast(x|\theta,Z)\psi(x|\theta,Z)
\right\}
.
\label{SE9}
\end{equation}
Substituting
\begin{equation}
\psi(x|\theta,Z)=\sqrt{P(x|\theta,Z)}e^{iS(x)\sqrt{\lambda}/2}
\label{psipsi}
\end{equation}
yields Eq.~(\ref{SE8}).

On the other hand, from a standard calculation using the variation
$\psi^\ast(x|\theta,Z) \rightarrow \psi^\ast(x|\theta,Z)+\delta\psi^\ast(x|\theta,Z)$,
it follows that the extrema of Eq.~(\ref{SE9}) are given
by the solutions of the linear eigenvalue problem,
\begin{equation}
-\frac{\partial^2 \psi(x|\theta,Z)}{\partial x^2}
+\frac{m\lambda}{2}\left[V(x)-E\right]\psi(x|\theta,Z)=0
,
\label{SE11}
\end{equation}
which is nothing but the time-independent Schr\"odinger equation
with $\lambda=4/\hbar^2$.
Planck's constant $\hbar$ enters here because of dimensional reasons
(see also Section~\ref{SEz}) and it sets the energy scale of
experiments which belong to category 3a.
As Eq.~(\ref{SE11}) is a linear second-order partial differential
equation, in practice computing its solution requires the
specification of two boundary conditions on $\psi(x|\theta,Z)$.

The equivalence between Eq.~(\ref{SE8}) and Eq.~(\ref{SE9})
was established by representing the two real-valued functions $P(x|\theta,Z)$
and $S(x)$ by the {\it complex}-valued function $\psi(x)$~\cite{MADE27}.
Given a solution $\psi(x|\theta,Z)$ of Eq.~(\ref{SE11}),
it follows that $P(x|\theta,Z)=|\psi(x|\theta,Z)|^2$
and $S(x)=-i\ln (\psi(x|\theta,Z)/\psi^\ast(x|\theta,Z))$ whenever $\psi(x|\theta,Z)\not=0$.
Clearly, because of the complex logarithm, the mapping from $\psi(x|\theta,Z)$
to the real-valued function $S(x)$ is not one-to-one~\cite{WALL94,VOLO09}.
In the hydrodynamic form of quantum theory~\cite{MADE27},
the ambiguity that ensues has implications for the interpretation
of the gradient of $S(x)$ as a velocity field~\cite{WALL94,VOLO09}.
As pointed out by Novikov, similar ambiguities appear in classical mechanics proper
if the local equations of motion (Hamilton equations) are not sufficient to characterize
the system completely and the global structure of the phase space
has to be taken into consideration~\cite{NOVI82}.
However, for the present purpose, this ambiguity has no effect on the minimization
of $F(\theta)$ because Eq.~(\ref{SE8}) does not change
if we add to $S(x)$ a real number which does not depend on $x$ or, equivalently,
if we multiply $\psi(x|\theta,Z)$ by a global phase factor.

For the experiment considered in the current (and next) section
the equation describing the experiment is Eq.~(\ref{SE8}), not the Schr\"odinger equation Eq.~(\ref{SE11}),
and the wavefunction $\psi(x|\theta,Z)$ is merely a vehicle to solve a set of nonlinear equations
through the solution of a linear eigenvalue problem.
This is logically consistent with the logical-inference treatment of the EPRB and Stern-Gerlach
experiments (see Sections~\ref{EPRB} and \ref{SG})
where there is no need to introduce a ``wavefunction'' $\psi(x|\theta,Z)$ to find closed form solutions.

In the case that the solutions of Eq.~(\ref{SE11})
are real-valued, we have $S(x) = 0 \bmod{2\pi}$.
Hence, it would be sufficient that $\psi(x|\theta,Z)$ is a {\it real}-valued function.
On the other hand, it is a simple matter to repeat the derivation and show that
minimization of the Fisher information with the constraint that
on average, the HJE of a particle in an electromagnetic field
should hold leads to the corresponding time-independent Schr\"odinger equation (see also Section~\ref{TDSE}).
Then, in general, it is necessary to introduce a
complex-valued function $\psi(x|\theta,Z)$
to linearize the minimization problem.

\subsection{Discussion}\label{SEd}

Starting from the assumptions that the experiment belongs to category 3a
and averages of the observed data complies with Newtonian mechanics,
application of logical inference straightforwardly leads to the
time-independent Schr\"odinger equation Eq.~(\ref{SE11}).
The key step in this derivation, which in essence
is the same as in Sections~\ref{EPRB} and \ref{SG},
is to express the robustness of the observed data (distribution of frequencies of the events)
with respect to small variations in the unknown position of the particle,
taking into account the inference that we draw on the basis
of the observed data, namely that on average there is agreement
with Newtonian mechanics.

Of course, a priori there is no good reason to assume that
on average there is agreement with Newtonian mechanics.
The only reason to do so here is that only then we
recover the time-independent Schr\"odinger equation.
In other words, the time-independent Schr\"odinger equation describes
the collective of repeated experiments of category 3a subject
to the condition that the averaged observations
comply with Newtonian mechanics.
The question what kind of equations are obtained
by assuming a different kind of ``mechanics'' is out
of the scope of the present paper.

It is very important to emphasize that from the logical-inference viewpoint the superposition principle, that is,
the linearity of the Schr\"{o}dinger equation, is not fundamental but follows
from the fact that in classical mechanics the kinetic energy is quadratic in the velocities and, thus,
in the momenta. Only in this case the substitution of Eq.~(\ref{psipsi}) reduces the nonlinear minimization
problem to a linear equation.
This raises the question what to do with different types of classical mechanics, such as relativistic mechanics.
It is well known, that {\it relativistic} quantum mechanics cannot be {\it mechanics}, it
can only be a {\it field theory}~\cite{LAND31,PAUL34,BERE71},
the argument being that any attempt to measure the coordinate of a particle with
the accuracy better than its Compton wavelength unavoidably leads to the creation of particle-antiparticle pairs.
We leave the challenging problem of
extending the present work to the relativistic domain for future research.

A comment on the identification $\lambda=4/\hbar^2$ is in order.
Clearly, from a dimensional analysis of Eq.~(\ref{SE11}), $\lambda$ has to be a parameter
with the dimension of $\hbar^{-2}$ but there is no a-priori reason
why we must have $\lambda=4/\hbar^2$.
However, comparing the results of a numerical calculation based on Eq.~(\ref{SE11}) with specific experimental results for
the spectra of atoms etc., we are forced to choose $\lambda=4/\hbar^2$.
It is worth mentioning here that the logical-inference derivation of the canonical ensemble of statistical mechanics~\cite{JAYN57a,JAYN57b}
employs the same reasoning
to relate the inverse temperature $\beta=1/k_BT$ to the average thermal energy.
At this point it should be mentioned that recent work has shown that $\hbar$
may be eliminated from the basic equations of (low-energy) physics
by a re-definition of the units of mass, time, etc.~\cite{VOLO10,RALS13a}.

A very important point, which renders our treatment very different
from other statistical formulations of
quantum theory~\cite{FRIE89,VSTO95,REGI98,HALL00,LUO02,FRIE04,KHRE09,KAPS10,KHRE11,KHRE11a,SKAL11,KAPS11,KLEI12,KLEI12a,FLEG12}
is that the unknown position of the particle $\theta$ never appears in the solution of the problem.
It appears as a condition on the i-probs but it has no effect on the functional dependence
of the i-probs on the relevant, observable coordinate $x$.

From quantum theory we know that Eq.~(\ref{SE11}) usually has
more than one solution, the minimum of Eq.~(\ref{SE9}) corresponding
to the quantum state with the lowest energy and the others being excited states.
The latter correspond to extrema of Eq.~(\ref{SE8}) with values of $F(\theta)$
that are larger than the minimum value of $F(\theta)$.

From Eq.~(\ref{SE12}) it also follows that the excited quantum states describe
experiments which are not the ``most'' robust against
small changes of $\theta$ but nevertheless have
the property that, to first order,
the ``quality'' of the results (i.e. averages, etc.)
does not depend on the particular value of $\theta$.

If we were to follow the tradition of conventional statistics,
we would introduce, for instance, an estimator $\widehat\theta(x)$ for $\theta$
and assume that the expectation value of this estimator relates to the ``true'' position of the particle.
As we know from the early days of the development of quantum theory~\cite{HEIS25}
trying to interpret such estimators as objective properties
of the particle creates seemingly endless possibilities for different interpretations,
paradoxes, and mysteries~\cite{HOME97}.
From the viewpoint of logical inference,
$\theta$ was and remains unknown and any attempt to interpret
the function $\psi(x)$ seems superfluous; $\psi(x)$ is just
an extremely useful vehicle to compute the numerical values of the i-probs $P(x|\theta,Z)$.

\subsection{Historical note}\label{SEz}

It is of interest to repeat here the first few steps in
Schr\"odinger's first paper on his equation~\cite{SCHR26a}.
For simplicity, we consider a particle moving on a line only.
Schr\"odinger starts from the time-independent HJE.
\begin{equation}
H\left(x,\frac{\partial S(x)}{\partial x}\right) =E
,
\label{HIS0}
\end{equation}
where
\begin{equation}
H\left(x,\frac{\partial S(x)}{\partial x}\right)=\frac{1}{2m}\left(\frac{\partial S(x)}{\partial x}\right)^2 + V(x)
,
\label{HIS0b}
\end{equation}
is the Hamiltonian of the classical, Newtonian particle.
Then, in Eq.~(\ref{HIS0}) he substitutes
\begin{equation}
S(x)=K \ln \psi(x)
,
\label{HIS0a}
\end{equation}
where $\psi(x)$ is assumed to be a real single-valued function
of $x$ and $K$ is a constant with the dimension of action
and obtains
\begin{equation}
H\left(x,\frac{K}{\psi(x)}\frac{\partial \psi(x)}{\partial x} \right) =E
.
\label{HIS1}
\end{equation}
Then, Schr\"odinger observes that one can rewrite
Eq.~(\ref{HIS1}) as a quadratic form in $\psi(x)$, namely
\begin{equation}
\frac{K^2}{2m}\left(\frac{\partial \psi(x)}{\partial x}\right)^2 + [V(x)-E]\psi^2(x) = 0
.
\label{HIS2}
\end{equation}
Of course, solving Eq.~(\ref{HIS2}) for
$\psi(x)$ does not bring anything new.
Therefore, Schr\"odinger postulates that
instead of solving Eq.~(\ref{HIS2}),
one should search for the extrema of the functional
\begin{equation}
Q=\int_{-\infty}^{+\infty} dx\;
\left[
\frac{K^2}{2m}\left(\frac{\partial \psi(x)}{\partial x}\right)^2 + [V(x)-E]\psi^2(x)
\right]
,
\label{HIS3}
\end{equation}
knowing that the formal solution of this variational problem
leads to an eigenvalue problem.
He then continues to show that by using
the classical Hamiltonian for the Kepler problem,
the solution of the eigenvalue problem yields the spectrum
of the hydrogen atom.

It is quite remarkable that in his next publication on the subject~\cite{SCHR26b},
Schr\"odinger calls both the ansatz Eq.~(\ref{HIS0a}) and
the transition from Eq.~(\ref{HIS2}) to Eq.~(\ref{HIS3})
incomprehensible (``unverst\"andlich'') and
then goes on to motivate his equation using
the analogy with optics.
As shown by our derivation of Eq.~(\ref{SE11}),
which on choosing $\lambda=4K^{-2}$ is the same as
Eq.~(\ref{HIS2}), from the viewpoint of logical inference applied
to experiments of category 3a, there is nothing
incomprehensible to Eq.~(\ref{HIS3}).

\section{Time-dependent Schr\"odinger equation}\label{TDSE}

Extending the reasoning which yields
the time-independent Schr\"odinger equation to the
time-dependent, multidimensional case does not
require new concepts but simply replacing
the position on the line by a vector
in 3D space and adding time labels does not suffice.
Therefore, in what follows, we focus
on those aspects which are absent
in the examples treated in Sections~\ref{EPRB}--\ref{SE}.

\subsection{Experiment}\label{TDSEa}

We consider $N$ repetitions of a thought experiment on
a particle moving in a $d$-dimensional hypercube $\bm\Omega$
of linear extent $[-L,L]$,
relative to a fixed reference frame.
Here and in the following $d$ is a positive integer.
A source emits a signal at discrete times labeled by the integer $\tau=1,\ldots,M$.
It is assumed that for each repetition, the particle
is at the unknown position $\bm\theta_\tau\in \bm\Omega$.
As the particle receives the signal, it responds by emitting
another signal which is recorded by an array of detectors.
For each signal emitted by a particle
the data recorded by the detector system
is used to determine the position $\bm{j}_{n,\tau}$
of a voxel of linear extent
$[-\Delta,\Delta]$ in the $d$-dimensional space $\bm\Omega$.
The dimension of the voxels determines the spatial resolution of the detection system.
As in Section~\ref{SE}, in a later stage, we will let $\Delta\rightarrow0$
to solve the problem analytically.

The result of $N$ repetitions of the experiment yields the data set
\begin{equation}
\Upsilon=\{\bm{j}_{n,\tau}| \bm{j}_{n,\tau}\in [-L^d,L^d];\; n=1,\ldots,N;\; \tau=1,\ldots,M \}
,
\label{TDSE0}
\end{equation}
or, denoting the total counts of voxels $\bm{j}$ at time $\tau$ by $0\le k_{\bm{j},\tau}\le N$,
the experiment produces the data set
\begin{eqnarray}
{\cal D}&=&\Bigl\{ k_{\bm{j},\tau}\Bigl| \tau=1,\ldots,M\;; N=\sum_{\bm{j}\in [-L^d,L^d]} k_{\bm{j},\tau} \Bigr\}
.
\label{TDSE1}
\end{eqnarray}
\subsection{Inference-probability of the data produced by the experiment}\label{TDSEb}

In analogy with the procedure followed in the previous sections,
we introduce the i-prob $P(\bm{j}|\bm\theta,\tau,Z)$
to describe the relation between the unknown location $\bm\theta$
and the location $\bm{j}$ of the voxel determined by the detector system
at discrete time $\tau$.
Except for the unknown location $\bm\theta$, all
other experimental conditions are represented by $Z$ and are assumed
to be fixed and identical for all experiments.
Note that unlike in the case of parameter estimation,
in the case at hand both $P(\bm{j}|\bm{\theta},\tau,Z)$ and the parameter $\bm\theta$ are unknown.
As in all examples treated so far, the key question is what the requirement of
reproducibility tells us about the i-prob $P(j|\bm{\theta},\tau,Z)$ as a function of $\bm{\theta}$.

The following assumptions are essentially the same as those of
Sections~\ref{EPRBb},~\ref{SGb}, and~\ref{SEb}.
\begin{enumerate}[1.]
\item
It is assumed that each repetition of the experiment represents an identical event of which
the outcome is logically independent of any other such event.
By application of the product rule, the consequence of this assumption is that
\begin{equation}
P({\cal D}|\bm{\theta}_1,\ldots,\bm{\theta}_M,N,Z)
= N!\prod_{\tau=1}^M\prod_{\bm{j}\in [-L^d,L^d]} \frac{P(\bm{j}|\bm\theta_\tau,\tau,Z)^{k_{\bm{j},\tau}}}{k_{\bm{j},\tau}!}
.
\label{TDSE2}
\end{equation}
\item
As in Section~~\ref{SEb}, we assume that space is homogeneous.
This implies that the i-prob has the property
\begin{equation}
P(\bm{j}|\bm\theta,Z)=P(\bm{j}+\bm\zeta|\bm\theta+\bm\zeta,Z)
,
\label{TDSE4}
\end{equation}
where $\bm\zeta$ is an arbitrary vector in $d$-dimensional space.
\end{enumerate}

\subsection{Condition for reproducibility and robustness}\label{TDSEc}

In Sections~\ref{EPRBb},~\ref{SGb} and~\ref{SEb}
the variable condition $\theta$ is a scalar
variable whereas in the present case, $\bm\theta$
denotes a collection of $d$ scalars.
This has some impact on the expression for the evidence.
Repeating the steps that took us from Eq.~(\ref{robu2})
to Eq.~(\ref{robu5}), we find that
\begin{equation}
\mathrm{Ev}=
\sum_{\bm{j},\tau}
\sum_{i,i'=1}^d
\frac{\epsilon_{i,\tau}\epsilon_{i',\tau}}{P(\bm{j}|\bm\theta_\tau,\tau,Z)}
\frac{\partial P(\bm{j}|\bm\theta_\tau,\tau,Z)}{\partial\theta_{i}}
\frac{\partial P(\bm{j}|\bm\theta_\tau,\tau,Z)}{\partial\theta_{i'}}
,
\label{TDSE5}
\end{equation}
where we have dropped the irrelevant prefactor $-N/2$ and
omitted from the summation sign the range of $\tau$ and $\bm{j}$ (see Eq.~(\ref{TDSE2}))
and the terms of third and higher order in the $\epsilon$'s.

The condition for reproducibility applied to Eq.~(\ref{TDSE5})
requires that we minimize $\mathrm{Ev}$ (which is non-negative, see Eq.~(\ref{TDSE6a})).
A minor problem thereby is that the $\epsilon_i$'s are arbitrary (but small)
but we can get around this problem by noting that
\begin{equation}
\mathrm{Ev}=
\sum_{\bm{j},\tau}
\left(\sum_{i=1}^d
\frac{\epsilon_i,\tau}{\sqrt{P(\bm{j}|\bm\theta_\tau,\tau,Z)}}
\frac{\partial P(\bm{j}|\bm\theta_\tau,\tau,Z)}{\partial\theta_{i}}
\right)^2
\ge 0
,
\label{TDSE6a}
\end{equation}
and, by using the Cauchy-Schwarz inequality, that
\begin{eqnarray}
\mathrm{Ev}&\le&
\sum_{\bm{j},\tau}
\left(
\sum_{i=1}^d \epsilon_{i,\tau}^2
\right)
\left(
\sum_{i=1}^d
\frac{1}{P(\bm{j}|\bm\theta_\tau,\tau,Z)}
\left(\frac{\partial P(\bm{j}|\bm\theta_\tau,\tau,Z)}{\partial\theta_{i}}\right)^2
\right)
\nonumber \\
&\le&d\widehat\epsilon^2
\sum_{\bm{j},\tau}
\sum_{i=1}^d
\frac{1}{P(\bm{j}|\bm\theta_\tau,\tau,Z)}
\left(\frac{\partial P(\bm{j}|\bm\theta_\tau,\tau,Z)}{\partial\theta_{i}}\right)^2
,
\label{TDSE6}
\end{eqnarray}
where $\widehat\epsilon^2=\max_{i,\tau}\epsilon_{i,\tau}^2$.
From Eq.~(\ref{TDSE6}) it follows that as the
$\epsilon_i$'s are arbitrary (but small), minimizing the rightmost factor in Eq.~(\ref{TDSE6})
is the best we can do to make sure that $\mathrm{Ev}$ is as small as possible.
Therefore, we find that in order to realize the condition for reproducibility
we have to minimize the Fisher information
\begin{equation}
I_F=
\sum_{\bm{j},\tau}
\sum_{i=1}^d
\frac{1}{P(\bm{j}|\bm\theta_\tau,\tau,Z)}
\left(\frac{\partial P(\bm{j}|\bm\theta_\tau,\tau,Z)}{\partial\theta_{i}}
\right)^2
,
\label{TDSE8}
\end{equation}
subject to additional constraints that we impose (see below).

As before, to make contact with the
Schr\"odinger equation which is formulated in continuum space-time,
it is necessary to replace sums over space-time coordinates by integrals.
Invoking translational invariance (see Section~\ref{TDSEb}),
we have
\begin{equation}
I_F=
\int d\bm{x} \int dt\;
\sum_{i=1}^d
\frac{1}{P(\bm{x}|\bm\theta(t),t,Z)}
\left(\frac{\partial P(\bm{x}|\bm\theta(t),t,Z)}{\partial x_i}
\right)^2
,
\label{TDSE9}
\end{equation}
where $\bm{x}=(x_1,\ldots,x_d)$.

We include the knowledge that
the particle moves in a time-dependent electromagnetic field and time-dependent potential
by repeating the steps of Section~\ref{SE} that lead us from Eq.~(\ref{SE7a}) to Eq.~(\ref{SE8}),
that is we start from the classical HJE and then account for the uncertainties about the events.

According to classical mechanics, the motion of a particle with mass $m$
in a time-dependent electromagnetic field and time-dependent potential is governed by
the time-dependent HJE
\begin{equation}
\frac{\partial S(\bm{\theta},t)}{\partial t}+\frac{1}{2m}\left(\bm\nabla S(\bm{\theta},t)-\frac{q}{c}\bm{A}(\bm{\theta},t)\right)^2
+V(\bm{\theta},t)=0
,
\label{TDSE9b}
\end{equation}
where $q$ denotes the electrical charge of the particle,
$c$ is the velocity of light in vacuum, $\bm{A}(\bm{x},t)$
represents the vector potential and the electrical potential and all potentials of non-electromagnetic origin
are collectively denoted by $V(\bm{x},t)$.

Using the same argument as the one in Section~\ref{SE},
if there is uncertainty about the position $\bm\theta$
but not about the individual event (category 2 experiment),
the simplest equation for $S(\bm{x},t)$ which
reduces to Eq.~(\ref{TDSE9b}) in the limit that there is no uncertainty reads
\begin{equation}
\int_{-\infty}^\infty d\bm{x}\;
\left[
\frac{\partial S(\bm{x},t)}{\partial t}+\frac{1}{2m}\left(\bm\nabla S(\bm{x},t)-\frac{q}{c}\bm{A}(\bm{x},t)\right)^2
+V(\bm{x},t)
\right]
P(\bm{x}|\bm\theta(t),t,Z)
=0
,
\label{TDSE9c}
\end{equation}
for each value of $t$.
If there is uncertainty
about both the individual event and the conditions (category 3 experiment),
the i-prob $P(\bm{x}|\bm\theta(t),t,Z)$ is unknown but can be determined
by requiring that the frequency distributions of the observed events are robust (category 3a experiment).
Note that no assumption about the {\it unknown} position $\bm\theta$ of the particle has been or will be made
and that this line of reasoning, which is reminiscent of Ehrenfest's theorem~\cite{EHRE27},
does not determine $P(\bm{x}|\bm\theta(t),t,Z)$ but merely provides a constraint on it.

Minimizing the Fisher information Eq.~(\ref{TDSE9}) with the
constraint Eq.~(\ref{TDSE9c}) amounts to minimizing the functional
%
\begin{eqnarray}
F&=&
\int d\bm{x} \int dt\;
\sum_{i=1}^d
\left\{
\frac{1}{P(\bm{x}|\bm\theta(t),t,Z)}
\left(\frac{\partial P(\bm{x}|\bm\theta(t),t,Z)}{\partial x_i}\right)^2
\right.
\nonumber \\
&&\hskip 10pt
\left.
+\lambda
\left[
\frac{\partial S(\bm{x},t)}{\partial t}
+
\frac{1}{2m}
\left(\frac{\partial S(\bm{x},t)}{\partial x_i}-\frac{q}{c}\bm{A}(\bm{x},t)\right)^2
+V(\bm{x},t)
\right]P(\bm{x}|\bm\theta(t),t,Z)
\right\}
,
\label{TDSE10}
\end{eqnarray}
where $\lambda$ is a Lagrange parameter
and the normalization of
$P(\bm{x}|\bm\theta(t),t,Z)$ can be taken
care of by exploiting the invariance of
the extrema of Eq.~(\ref{TDSE10}) with respect
to the rescaling transformation  $P(\bm{x}|\bm\theta(t),t,Z)\rightarrow\alpha P(\bm{x}|\bm\theta(t),t,Z)$.
Note that the integrand of Eq.~(\ref{TDSE10}) is invariant for the {\it gauge transformation}
$\bm{A}(\bm{x},t)\rightarrow\bm{A}(\bm{x},t)+(q/c)\bm\nabla\chi(x,t)$,
$V(\bm{x},t)\rightarrow V(\bm{x},t)-(q/c)\partial\chi(x,t)/\partial t$,
and
$S(\bm{x},t)\rightarrow S(\bm{x},t)+(q/c)\chi(x,t)$
where $\chi(x,t)$ is an arbitrary scalar function~\cite{KARL13}.

As in Section~\ref{SE}, it follows that
finding the extrema of the functional Eq.~(\ref{TDSE10}) is tantamount
to solving the time-dependent Schr\"odinger equation (TDSE).
Applying the standard variational argument, it follows that the
solutions of the TDSE
\begin{equation}
i\hbar \frac{\partial \psi(\bm{x}|\bm\theta(t),t,Z)}{\partial t}
=\left[
-\frac{\hbar^2}{2m}\sum_{j=1}^d\left(
\frac{\partial }{\partial x_j}
-\frac{iq}{\hbar c}\bm{A}(\bm{x},t)\right)^2
+V(x,t)
\right]
\psi(\bm{x}|\bm\theta(t),t,Z)
,
\label{TDSE9a}
\end{equation}
are the extrema of the functional
\begin{eqnarray}
Q=
2\int d\bm{x} \int dt\;
&\bigg\{&
mi\sqrt{\lambda}\left[
\psi(\bm{x}|\bm\theta(t),t,Z)\frac{\partial \psi^\ast(\bm{x}|\bm\theta(t),t,Z)}{\partial t}
-
\psi^\ast(\bm{x}|\bm\theta(t),t,Z)\frac{\partial \psi(\bm{x}|\bm\theta(t),t,Z)}{\partial t}
\right]
\nonumber \\
&&+
2
\sum_{j=1}^d
\left(
\frac{\partial \psi^\ast(\bm{x}|\bm\theta(t),t,Z)}{\partial x_j}
+\frac{iq\sqrt{\lambda}}{2 c}{A}_j(\bm{x},t)\psi^\ast(\bm{x}|\bm\theta(t),t,Z)
\right)
\nonumber \\
&&
\hskip 1cm
\times
\left(
\frac{\partial \psi(\bm{x}|\bm\theta(t),t,Z)}{\partial x_j}
-\frac{iq\sqrt{\lambda}}{2 c}{A}_j(\bm{x},t)\psi(\bm{x}|\bm\theta(t),t,Z)
\right)
\nonumber \\
&&+
m\lambda V(\bm{x},t)\psi^\ast(\bm{x}|\bm\theta(t),t,Z)\psi(\bm{x}|\bm\theta(t),t,Z)
\bigg\}
,
\label{TDSE11}
\end{eqnarray}
if $\lambda=4/\hbar^2$. The equivalence of
Eq.~(\ref{TDSE10}) and Eq.~(\ref{TDSE11}) follows by substituting
$\psi(\bm{x}|\bm\theta(t),t,Z)=\sqrt{P(\bm{x}|\bm{\theta}(t),t,Z)}e^{iS(\bm{x},t)\sqrt{\lambda}/2}$.
Note that the solutions of Eq.~(\ref{TDSE9a}) do not depend
on the unknown position $\bm{\theta}(t)$, as it should be.
As in the case of the SE (see Section~\ref{SE}), it follows that the solutions of the variational problem
comply with the constraint (C3) (see Section~\ref{EPRBc}).

The functional Eq.~(\ref{TDSE11}) inherits from Eq.~(\ref{TDSE10}) the
invariance under gauge transformations. Specifically, it is
easy to show that the integrand in Eq.~(\ref{TDSE11}) does not change
by substituting
$\bm{A}(\bm{x},t)\rightarrow\bm{A}(\bm{x},t)+(q\sqrt{\lambda}/2c)\bm\nabla\chi(x,t)$,
$V(\bm{x},t)\rightarrow V(\bm{x},t)-(q/c) \partial\chi(x,t)/\partial t$,
and
$\psi(\bm{x}|\bm\theta(t),t,Z)\rightarrow \psi(\bm{x}|\bm\theta(t),t,Z) \exp[iq\sqrt{\lambda}\chi(x,t)/2c]$
where $\chi(x,t)$ is an arbitrary scalar function.

Instead of solving the nonlinear differential equation that follows
from extremizing Eq.~(\ref{TDSE10}), it is usually more expedient
to solve the linear partial differential equation Eq.~(\ref{TDSE9a}).
Of course, in practice
we need to specify $\psi(\bm{x}|\bm\theta(t),t,Z)$ at $t=0$
in order to solve the initial-value problem Eq.~(\ref{TDSE9a}).
Unlike in the time-independent case (see Section~\ref{SE})
where we may have solutions for which $S(x) = 0 \bmod{2\pi}$,
in the general case, the equivalence between Eq.~(\ref{TDSE10}) and Eq.~(\ref{TDSE11})
cannot be established unless we allow $\psi(\bm{x}|\bm\theta(t),t,Z)$
to be complex-valued.
In general, minimizing Eq.~(\ref{TDSE10}) yields solutions
for the two real-valued functions $P(\bm{x}|\bm{\theta}(t),t,Z)$ and $S(\bm{x},t)$
and although we can represent these two functions in a variety of
ways, the complex-valued representation
in terms of $\psi(\bm{x}|\bm\theta(t),t,Z)$
offers the for computational reasons very important
advantage that it transforms a nonlinear optimization problem
into a linear one.

The equivalence of Eq.~(\ref{TDSE10}) and Eq.~(\ref{TDSE11})
allows us to determine, from the solutions of the TDSE, the i-probs
$P(\bm{x}|\bm{\theta}(t),t,Z)$ which yield the most likely
and most reproducible data, collected in the
experiment described in Section~\ref{TDSEa}.
Put differently, through the TDSE,
quantum theory describes an experiment which yields data
that is the most robust with respect to small variations of the external conditions
(the unknown positions of the particle)
under which the experiment is being performed.

\subsection{Discussion}\label{TDSEd}

In essence, all the points that were mentioned in the discussions
in Sections~\ref{EPRB} -- \ref{SE} also hold
for the time-dependent case.
Of course, one should replace
for instance ``time-independent'' by ``time-dependent'',
but otherwise there are no significant conceptual changes.

Having shown how basic results of quantum theory follow from the application
of logical inference to experiments of category 3a,
it is appropriate to compare our approach to the considerable body of
work~\cite{FRIE89,VSTO95,REGI98,HALL00,LUO02,FRIE04,KHRE09,KAPS10,KHRE11,KHRE11a,SKAL11,KAPS11,KLEI12,KLEI12a,FLEG12}
which shows that quantum theory can be cast into a ``classical'' statistical theory.

Central in the derivations presented in the present paper is the appearance of the Fisher information.
Therefore, it is instructive to
compare the methodology adopted in the present paper with the ones of
earlier works~\cite{FRIE89,VSTO95,REGI98,HALL00,LUO02,FRIE04,KAPS10,SKAL11,KAPS11,KLEI12,KLEI12a,FLEG12}
in which it is de facto {\it postulated} that the Fisher information
is the basic expression from which the equations of theoretical physics can be derived.

To the best of our knowledge, the idea of postulating that
the Fisher information is the starting point for deriving the time-independent Schr\"odinger equation
appeared for the first time in a paper by Frieden~\cite{FRIE89}
who also showed that the Heisenberg-Robertson inequalities,
often regarded as a landmark of quantum mechanics,
directly follow from the Cram\'er-Rao inequality~\cite{FRIE89,HALL00,LUO02,FRIE04,KAPS10,SKAL11,KAPS11,KLEI12a},
a standard result in classical statistics~\cite{FRIE04,TREE68}.

The expressions of functionals akin Eq.~(\ref{TDSE10}) are justified
using arguments from estimation theory, and concepts such as intrinsic fluctuations
and ``smart measurements''.
Thereby, it seems essential that the difference between the parameter to be estimated (e.g. $\bm\theta(t)$
in our notation) and the measured quantity (e.g. $\bm{x}$)
may be interpreted as intrinsic fluctuations.

These earlier works~\cite{FRIE89,VSTO95,REGI98,HALL00,LUO02,FRIE04,KAPS10,SKAL11,KAPS11,KLEI12,KLEI12a,FLEG12}
have been instrumental for the development of our logical-inference approach.
But in contrast to these earlier works in which the Fisher information is postulated as the starting point,
in the logical-inference approach the Fisher information appears quite naturally as a result of
expressing the requirement that the experiment yields reproducible, robust results.
Thereby the notion of robustness used in the present paper refers to the effect of
small (systematic) changes of a parameter on the state of knowledge encoded in the i-probs and is conceptually
very different from the one introduced by Hall which expresses resilience with respect to noise~\cite{HALL00}.

We further illustrate the conceptual differences between the logical-inference approach
and the Fisher-information approach using two concrete examples.
First, consider Frieden's treatment of the EPRB experiment~\cite{FRIE04}
in which the angle $\theta$ between the two unit vectors corresponding to the directions
of the two routers that are controlled by the experimenter is
regarded as the variable-to-be-estimated.
From the point of view of laboratory experiments this seems to be a rather artificial starting point.
Indeed, we do not know of any real EPRB experiment which attempts to estimate this angle (which experimenters
consider to be known).
Moreover, mathematically we cannot even define the difference
between the observed event $\{x,y\}$ and the ``estimated'' angle $\theta$,
let alone that we can interpret this difference as a signature of intrinsic fluctuations.
Yet, as we have shown in Section~\ref{EPRB}, straightforward application of logical inference
to an experiment assumed to belong to category 3a, effortlessly yields the equations of the quantum
theoretical description for this experiment.
Next, as another example illustrating the conceptual differences, consider
Reginatto's derivation of the time-dependent Schr\"odinger equation~\cite{REGI98}, the algebra
of which has been our source of inspiration to connect Eq.~(\ref{TDSE10}) and Eq.~(\ref{TDSE9a}).
From the point of view of probability theory, the justification of Reginatto's derivation is problematic.
In Ref.~\cite{REGI98} the Fisher information (matrix) is introduced in the conventional manner~\cite{FRIE04},
namely as an information measure of estimating a parameter $\theta$
from the observed random variable $y=\theta+x$, $x$ representing the additive noise.
Then, in the next step, the noise $x$ is tactically taken as the position of the particle(s),
a remarkable reinterpretation of mathematical symbols.
Logical difficulties of this kind are absent in the logical inference approach simply
because it is not permitted to drop conditional dependences of the i-probs.

In short, the main conceptual difference with earlier works that start by postulating
expressions containing the Fisher information~\cite{FRIE89,HALL00,LUO02,FRIE04,KAPS10,SKAL11,KAPS11,KLEI12a},
is that in the logical inference approach the expression to be minimized is not postulated but
it is derived by assuming that the theory describes reproducible experiments in the most robust possible way.

\section{Conclusion}\label{CONC}\label{DISC}

We have shown that the basic equations of quantum
theory derive from logical inference applied
to experiments in which there is uncertainty about individual events
but for which the frequencies of events are reproducible and
most insensitive to small variations of the
unknown factors.

The derivations presented in Sections~\ref{EPRB}--\ref{TDSE}
demonstrate that logical inference, that is plausible reasoning,
applied to experiments which belong to
\begin{enumerate}[\bf {Category} 3a.]
\item
There is uncertainty about each event,
the conditions under which the experiment is carried out may be uncertain,
and the frequencies with which events are observed are reproducible
and robust against small changes in the conditions,
\end{enumerate}
yields two important, general results.

The first is the justification of the intuitive procedure to assign
to the i-probs the frequencies for the events to occur.
For {\it fixed} experimental conditions,
the usual argument for adopting this assignment is that it maximizes the i-prob
to observe these frequencies (see the \ref{APP1}).
On the other hand, it is quite natural to expect that under {\it variable} experimental conditions
it is the most robust, reproducible experiment which produces the most likely frequencies of events.
Obviously, the argument based on reproducibility
under {\it variable} experimental conditions
is more general as it contains the condition
of {\it fixed} experimental conditions as a special case.

The second, and most important for the purpose of recovering the quantum theoretical
description as an application of logical inference,
are equations that determine the functional dependence of the i-probs
on the condition that is considered to be variable.
Application of exactly the same procedure
to the Einstein-Podolsky-Rosen-Bohm experiment,
the Stern-Gerlach experiment,
and experiments on a particle in a potential
demonstrate that the equations  known from the quantum theoretical
description of these experiments follow in a straightforward manner
without invoking concepts of quantum theory.

The key point in the derivation of
the quantum theoretical description is to express
precisely and unambiguously, using the mathematical
framework of plausible reasoning~\cite{COX46,COX61,TRIB69,SMIT89,JAYN03}, the
essential features of experiments belonging to category 3a.
Adding the requirement that the experimental results are insensitive to small changes
of the conditions under which the experiment is carried out
yields equations that are known from quantum theory.
Furthermore, it also explains that if it is difficult
to engineer nanoscale devices which operate in a regime where the data is reproducible,
it is also difficult to perform these experiments such that the data complies with quantum theory.

In our logical inference derivation of the time-independent and time-dependent
Schr\"odinger equation we did not assume that the former can be deduced from the latter:
both emerge as descriptions of the data obtained from  different kinds of experiments.
Once both descriptions have been formulated in terms of quadratic forms,
the machinery of linear algebra brings out the equivalence of these two descriptions.
However, from a logical-inference viewpoint, there is no a-priori reason why this connection
should exist and therefore it is a logical, not physical, requirement that the logical inference
approach can be applied to the time-independent and time-dependent data without
assuming that there is a deeper relation between the two.
In a sense, the mathematical relationship that appears is a result of logical inference.
As mentioned in the introduction, this way of thinking is different from the deductive, reductionist approach.
For instance, it has recently been suggested that starting from real-valued Majorana-fermion equations,
one can derive the complex-valued Weyl equation, then reduce it to the Dirac equation from
which the time-dependent Schr\"odinger equation follows by taking the non-relativistic limit~\cite{VOLO14}.
In the reductionist approach, a description of the observed phenomena at low energy emerges
from an appropriate low-energy approximation of the underlying high-energy model~\cite{VOLO14}.
In contrast, in the logical inference approach, we take the point of view that a description of
our knowledge of the phenomena at a certain level is independent of the description at a more detailed level.
Of course, this implies that it should be possible to show that e.g. the Dirac and Klein--Gordon equation
can be obtained by logical inference applied to data collected in some (thought) experiment,
without making any reductionist detour.
Clearly, such a demonstration would be a very important step for establishing the usefulness of
the logical inference as a methodology to construct descriptions of observed phenomena.

The logical-inference methodology to derive the basic equations of quantum
theory has some implications for interpretational aspects of quantum theory.
First, although it supports Bohr's view expressed in quotes (1--3) of the introduction,
it does not support the Copenhagen interpretation (in any form)~\cite{HOME97}.
Indeed, the wavefunction Eq.~(\ref{psipsi}) merely appears to be a purely mathematical vehicle
to turn nonlinear differential equations into linear ones and it
seems difficult to attribute more meaning to such a vehicle other than mathematical.
On the other hand, there is no conflict with the statistical interpretation~\cite{BALL03,BALL70}
if we ignore the conceptual difference between i-probs and ``mathematical'' probabilities.
Second, it follows that quantum theory is a ``common sense'' description of the
vast class of experiments that belongs to category 3a.
Quantum theory definitely does not describe what is happening to a particle, say.
This follows most clearly from our derivation of the Schr\"odinger equation,
which shows that quantum theory does not provide {\it any}
insight into the motion of a particle but instead describes
all what can be {\it inferred} (within the framework of logical inference) from
or, using Bohr's words, {\it said} about the observed data,
in complete agreement with Bohr's view expressed in quotes (1--3) of the introduction.

The logical-inference derivation of the quantum theoretical description
does not, in any way, prohibit the construction of cause-and-effect
mechanisms that, when analyzed in the same manner as
in real experiments, create the {\it impression}
that the system behaves as prescribed by quantum theory~\cite{THOO97,THOO07,RAED05b}.
From Bohr's quote (1) reproduced in the introduction,
and as demonstrated in a mathematically rigorous manner
in the present paper, quantum theory is but an abstract description,
be it a very powerful one.
As mentioned in Section~\ref{sec3}, it is straightforward
to construct computer simulation models that mimic,
for all practical purposes almost perfectly, experiments that belong to
category 3a.
Work in this direction, for a review see Ref.~\cite{RAED12a},
has shown that it is indeed possible to build simulation models
which reproduce, on an event-by-event basis, (quantum) interference and entanglement
phenomena.

Summarizing: In line with Bohr's statement that
``Physics concerns what we can {\it say} about nature~\cite{PETE63}'',
the aim of physics is to provide a consistent description of
relations between certain classes of events.
Some of these relations express cause followed by an effect and others do not.
If there are uncertainties about the individual events
and the conditions under which the experiment is carried out,
situations may arise in which it becomes difficult or even
impossible to establish relations between individual events.
In the case that the frequencies of these events
are reproducible and robust, it may still be possible
to establish relations, not between the individual events,
but between the frequency distributions of the observed events.
As we have demonstrated, it is precisely under
these circumstances that the application
of logical inference to the (abstraction of) the experiment
yields the basic equations of quantum theory.
This then also explains the reason for the extraordinary descriptive power
of quantum theory: it is plausible reasoning, that is common sense, applied
to reproducible and robust experimental data.
The algebra of logical inference facilitates
this reasoning by means of a mathematically precise language which
is unambiguous and independent of the individual.

\section*{Acknowledgement}
We would like to thank Koen De Raedt, Karl Hess, Fengping Jin,  Andrei Khrennikov,
Thomas Lippert, Seiji Miyashita, Theo Nieuwenhuizen, Arkady Plotnitsky, and Grigori Volovik
for many stimulating discussions.

\appendix
\section{Maximum of the inference-probability}\label{APP1}
We consider an experiment with logically independent
outcomes $O_1,\ldots,O_m$ which is repeated $N$ times
under constant conditions represented by the proposition $Z$.
The i-prob that the outcome $O_k$ occurred $n_k$ times reads
\begin{equation}
P(n_1,\ldots,n_m|N,Z)=
\frac{N!}{n_1! \ldots n_m!} P(O_1|N,Z)^{n_1}\ldots P(O_m|N,Z)^{n_m}
.
\label{app0}
\end{equation}
Let us denote the set of values of $\{n_1,\ldots,n_m\}$ which maximizes $P(n_1,\ldots,n_m|N,Z)$ by
$\{n_1^\ast,\ldots,n_m^\ast\}$.
Then, we must have
\begin{equation}
\frac{P(n_1^\ast,\ldots,n_k^\ast,\ldots,n_m^\ast|N,Z)}{P(n_1^\ast+1,\ldots,n_k^\ast-1,\ldots,n_m^\ast|N,Z)}=
\frac{n_1^\ast+1}{n_k^\ast} \frac{P(O_k|N,Z)}{P(O_1|N,Z)}\ge1
,
\label{app2}
\end{equation}
or
\begin{equation}
n_k^\ast P(O_1|N,Z) \le (n_1^\ast+1) P(O_k|N,Z)
,
\label{app3}
\end{equation}
for all $2\le k\le m$.
Summing over all $k$ yields
\begin{eqnarray}
P(O_1|N,Z) \sum_{k=2}^m n_k^\ast &\le& (n_1^\ast+1)  \sum_{k=2}^m P(O_k|N,Z),
\nonumber \\
\noalign{or}
P(O_1|N,Z) (N-n_1^\ast) &\le& (n_1^\ast+1) (1-P(O_1|N,Z)),
\nonumber \\
\noalign{or}
P(O_1^\ast|N,Z) &\le& \frac{n_1^\ast+1}{N+1}
.
\label{app4}
\end{eqnarray}
Similarly, if we consider
\begin{equation}
\frac{P(n_1^\ast,\ldots,n_k^\ast,\ldots,n_m^\ast|N,Z)}{P(n_1^\ast-1,\ldots,n_k^\ast+1,\ldots,n_m^\ast|N,Z)}=
\frac{n_k^\ast+1}{n_1^\ast} \frac{P(O_1|N,Z)}{P(O_k|N,Z)}\ge1
,
\label{app5}
\end{equation}
we find
\begin{eqnarray}
P(O_1|N,Z) &\ge& \frac{n_1^\ast}{N+m-1}
.
\label{app6}
\end{eqnarray}
In the derivation that leads to Eqs.~(\ref{app4}) and (\ref{app6}),
our choice of the pair $(1,k)$ ($2\le k$) was arbitrary.
Repeating the derivation for $1\le j,k \le m$ with $k\not= j$ yields
\begin{equation}
\frac{n_j^\ast}{N+m-1} \le P(O_j|N,Z) \le \frac{n_j^\ast+1}{N+1} \quad,\quad 1 \le j \le m
,
\label{app7}
\end{equation}
or equivalently
\begin{equation}
P(O_j|N,Z)-\frac{1}{N+1} \le  \frac{n_j^\ast}{N+1} \le P(O_j|N,Z) \left(1+\frac{m-2}{N+1}\right)
,
\label{app8}
\end{equation}
for all $1 \le j \le m$.

For sufficiently large $N$, it follows from Eq.~(\ref{app7}) that for
an experiment with logically independent
outcomes $O_1,\ldots,O_m$ which is repeated $N$ times
under constant conditions represented by the proposition $Z$,
the assignment
\begin{equation}
P(O_j|N,Z)\leftarrow \frac{n_j}{N+1}  \quad,\quad 1 \le j \le m
,
\label{app9}
\end{equation}
maximizes the i-prob that $O_j$ occurs $n_j$ times for all $1\le j \le m$.

The derivation of the assignment Eq.~(\ref{app9}) justifies the intuitive procedure
to take as the numerical values of the i-probs, the frequencies of occurrences,
if the latter are known through actual measurement.

\section*{References}

\bibliography{../../../../all13,../qtextra}

\end{document}